\documentclass[preprint,12pt]{elsarticle}

\usepackage{graphicx}
\usepackage{amssymb}
\usepackage{amsthm}
\usepackage{mathtools}
\usepackage{lineno}
\usepackage{mathrsfs}
\usepackage{booktabs, caption, makecell}
\usepackage{threeparttable}
\usepackage{algorithm}
\usepackage{algcompatible}
\usepackage{lipsum}
\usepackage{appendix}
\usepackage{float}
\usepackage[colorlinks,linkcolor=blue]{hyperref}
\usepackage[nameinlink,capitalise]{cleveref}
\usepackage{calc}

\newcommand{\RNum}[1]{\uppercase\expandafter{\romannumeral #1\relax}}

\DeclareMathOperator*{\argmax}{arg \, max}

\theoremstyle{remark}
\newtheorem*{remark}{Remark}

\begin{document}

\begin{frontmatter}
\title{Dynamic mode decomposition of random pressure fields over bluff bodies}
\author[label1]{Xihaier Luo \corref{cor1}}
\ead{xluo1@nd.edu}
\author[label1]{Ahsan Kareem}
\ead{kareem@nd.edu}
\cortext[cor1]{Corresponding author. 156 Fitzpatrick Hall, Notre Dame, IN 46556, USA.}
\address[label1]{NatHaz Modeling Laboratory, University of Notre Dame,  Notre Dame, IN 46556, United States}

\begin{abstract}
    The proper orthogonal decomposition (POD), a modal decomposition technique, has been widely used for analyzing the spatial-temporal coherence of wind pressure fields, due to the complex nature of wind-structure interactions. The interpretation of characteristics and information of the extracted modes, however, is influenced by the POD algorithm, which tends to mix multiple temporal frequencies in a single mode. In this study, an operator-theoretic based decomposition approach referred to as the dynamic mode decomposition (DMD) is introduced to characterize aerodynamic pressure fields to offer an additional perspective. More specifically, the proposed approach can capture not only dominant spatial patterns but also identify each pattern with a specific frequency and a corresponding temporal evolution (growth/decay). To demonstrate the unique feature of the proposed decomposition algorithm, this study is directed towards the analysis of fluctuating pressures on bluff bodies immersed in boundary layer flows. In particular, aerodynamic pressure fields over a prism were analyzed using limited wind tunnel data and a comparison with the POD was carried out to observe commonalities and unique perspective of each decomposition. Quantification and interpretation of the decomposition results are provided. It is envisaged that DMD would supplement our understanding of the dynamics of pressure fields currently available via POD.
\end{abstract}


\end{frontmatter}

\section{Introduction}
\label{sec1}
Understanding bluff body aerodynamics is essential to estimate wind effects, for not only the performance-based design but also the control, operation, and maintenance of structures such as tall buildings \cite{cermak1976aerodynamics, kareem2008numerical}. Usually, aerodynamic characteristics such as fluctuating wind pressures on a bluff body are characterized as spatial-temporally varying random fields. Extensive research has been conducted to better extract dominant features embedded in these random pressure fields \cite{vickery1966fluctuating, lee1975effect, kareem1984pressure, tamura1999proper, chen2005proper, solari2007proper, carassale2011statistical}.

\begin{figure}[H]
\includegraphics[width=1.0\textwidth]{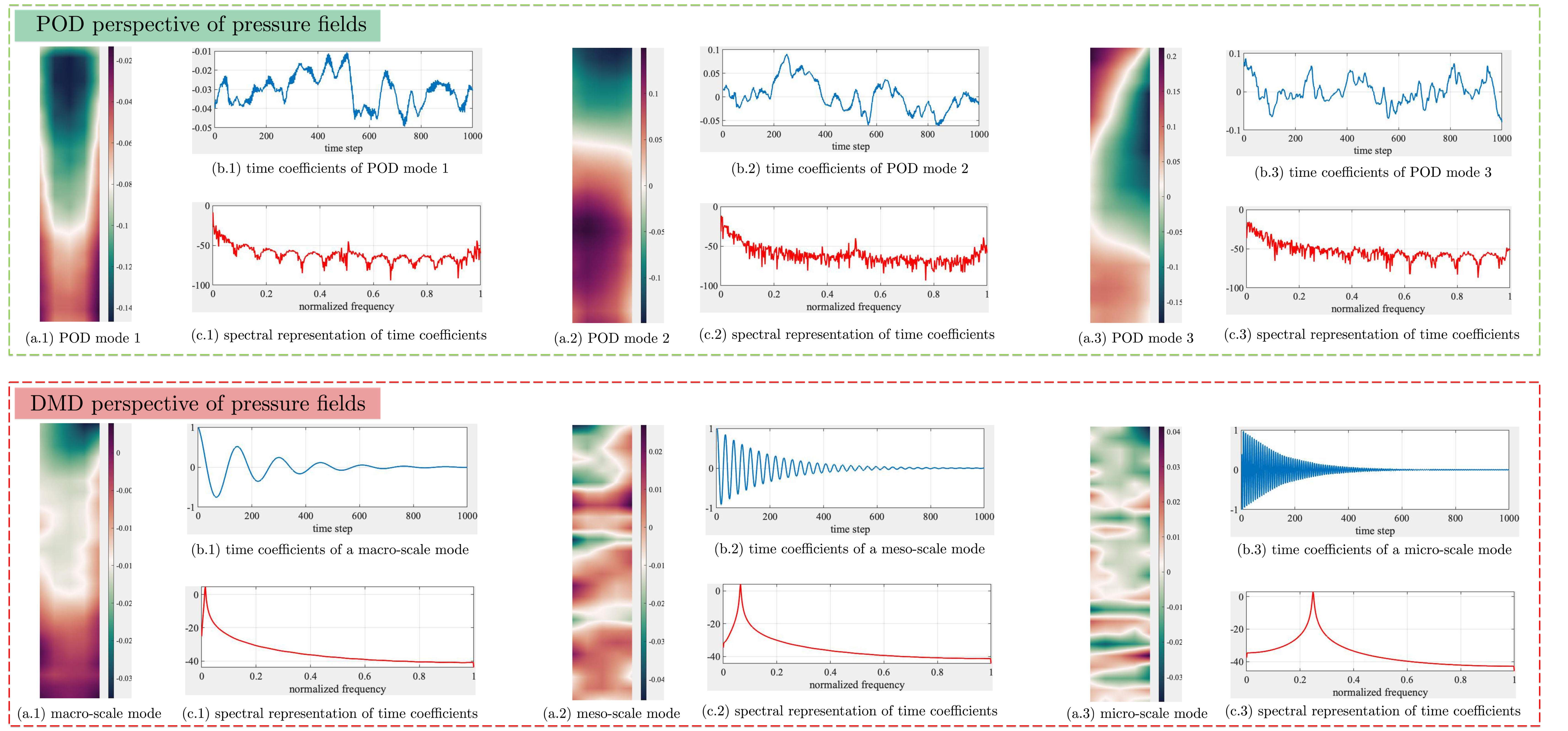}
\caption{Decomposition analysis for aerodynamic characteristics of a prism. The temporal behavior of POD modes contain a mix of frequencies while DMD can isolate each spatial pattern with a dominant frequency (See the spectrum of identified modes).} 
\label{fig: f1}
\end{figure}

Known under different names such as the Karhunen-Lo\`eve expansion, principal component analysis, etc, the proper orthogonal decomposition (POD) is one of the most widely used methods for analyzing random pressure fields. Specifically, POD analysis involves a procedure that transforms a multi-dimensional isotropic random field into a set of uncorrelated single-dimensional spatial modes, which are organized by a sequence of a continuous-valued random process \cite{jolliffe2011principal}. Because the identified spatial modes are orthogonal to each other, the POD method optimally evaluates the random pressure field in a $L_2$ norm sense. Consequently, the first few modes ranked on the basis of proportional energies are generally sufficient to recover the dynamics. Despite its wide range of applications, POD analysis is restricted to estimating second-order statistics, and its dimension-reduced representation is limited to a linear transformation range \cite{aubry1988dynamics, berkooz1993proper}. More importantly, the POD method by definition is unable to directly identify single-frequency dynamic coherent structures \cite{muld2012flow, zhang2014identification, taira2017modal, towne2018spectral}. Instead, it mixes a range of frequencies in a temporal behavior (See \cref{fig: f1}).

Note that fact the fluctuating wind pressure data, in the abstract sense, contains two elements: a set of states that describe the evolution of the pressure distribution in time, and an operator that defines the rule for that evolution. Therefore, integrating the evolution map into the decomposition process can potentially overcome the mixed multiple temporal frequencies issue encountered in POD analysis. This is the central idea of operator-theoretic modeling, especially the Koopman operator which lifts the dynamics from the state space to the space of \textit{observables} that are functions of the state variables \cite{koopman1931hamiltonian, budivsic2012applied, williams2015data, arbabi2017ergodic}. Following this principle, the nonlinear, finite-dimensional pressure dynamics can be mapped into a linear but infinite-dimensional Hilbert space.

However, solving such an infinite-dimensional representation can be very challenging. Different methods have been continually developed to obtain a numerical approximation of the Koopman operator. Dynamic mode decomposition (DMD), which is an equation-free method and was initially introduced by Schmid out of a need to identify spatio-temporal patterns buried in fluid flows, was found closely connected to the spectral analysis of the Koopman operator \cite{schmid2010dynamic, rowley2009spectral, chen2012variants, tu2013dynamic}. In particular, DMD decomposes a linearly dependent operator that is determined by a pair of time-shifted matrices to approximate the spectral properties associated with the Koopman operator. This results in a group of periodically oscillating eigenmodes. The resolution scheme is designed in an Arnoldi-algorithm like architecture, where the explicit formulation of evolution map can be saved, greatly improving the computational efficiency \cite{mezic2013analysis, kutz2016dynamic}.

In this study, we attempt to learn spatial-temporal pressure patterns utilizing operator-theoretic modeling method, i.e. the Koopman operator. The corresponding dynamic coherent structures will be resolved by means of an augmented dynamic mode decomposition algorithm. The augmentation centers on enriching a limited wind tunnel dataset, which is often the case in experimental research. Specifically, the manifold representing the state space is embedded into a higher dimensional Euclidean space with topological properties staying unchanged. This topologically equivalent representation is achieved by the implementation of Takens' Embedding Theorem \cite{takens1981detecting, le2017higher, takeishi2017learning}. In particular, delay coordinates containing informative content of the dynamical evolution are exploited. As a result, the augmented dataset is able to provide sufficient information for the operator-theoretic modeling of fluctuating wind pressures. We demonstrate the enhanced capability by applying the proposed operator-theoretic approach to learn the dynamics of random pressure fields over surfaces of a scaled model of a finite height prism. Compared to classic methods such as POD, learning results indicate the augmented DMD is able to isolate identified coherent structures with a specific frequency as well as a corresponding temporal growth/decay (See \cref{fig: f1}). Moreover, we connected the identified pressure patterns to the wind environmental conditions around the prism, explaining the pressure dynamics from a turbulence perspective \cite{tennekes1972first, leonard1975energy, sirovich1987turbulence, holmes2012turbulence}.

The remainder of this paper is organized as follows. \cref{sec2} presents the problem we are dealing with and offers a detailed discussion on the solution method. \cref{sec8} gives two mathematical examples to demonstrate the principles of different decomposition algorithms. \cref{sec3} describes the experimental setup of the comparative case study. Subsequently, \cref{sec4}, \cref{sec5}, and \cref{sec6} summarize the POD/DMD results including algorithm convergence, data sufficiency, and modal analysis. Finally, concluding remarks and an outlook on future work are provided in \cref{sec7}.

\section{Methodology}
\label{sec2}
In this work, we specially focus on analyzing the fluctuating wind pressures over a prism. Without loss of generality, the aerodynamic pressure field is modeled as a stationary and ergodic multivariate random process \cite{kareem2008numerical, lee1975effect, carassale2011statistical}:

\begin{equation}
\label{eq: sec2_1}
\frac{d}{dt} \left( \boldsymbol{x} \right) = f \left( \boldsymbol{x} \left( t \right), \boldsymbol{\theta} \right)
\end{equation}

where $\boldsymbol{x}$ is the system state denoting the pressure distribution of a prism surface, and it is contained in the state space $S \subset \mathbb{R}^{d}$. $f: S \rightarrow \mathbb{R}^{d}$ is a vector field that depends on the state $\boldsymbol{x}$, time $t$, and a set of parameters $\boldsymbol{\theta}$ controlling the wind-structure interaction. To apply this evolution representation to the discretized wind tunnel data, \cref{eq: sec2_1} is reformulated as:

\begin{equation}
\label{eq: sec2_2}
\boldsymbol{x} \left( t_k \right) = \boldsymbol{F} \left( \boldsymbol{x}_{k-1} \right) = \boldsymbol{x} \left( t_{k-1} \right) + \int_{t_{k-1}}^{t_k} f \left( \tau, \boldsymbol{x} \left( \tau \right) \right) d \tau
\end{equation}

where $\boldsymbol{F}: S \rightarrow S$ is the flow map, $k$ is the discrete time index, and $\tau$ is a sufficiently small number.

The objective is to provide physical insights concerning the aerodynamic characteristics of fluctuating wind pressures by analyzing limited measurements. However, the pressure data is highly nonlinear and exhibits multi-scale pattern in both space and time. To address this nonlinearity and complexity issue, an operator-theoretic approach is considered.

\subsection{The Koopman operator}
\label{sec21}
Consider the dynamical system given in \cref{eq: sec2_2}, let $g: S \rightarrow \mathbb{R}$ to be a real-valued \textit{observable} that is an element of an infinite-dimensional Hilbert space. The collection of a complete set of \textit{observables} constitutes a linear vector space. The Koopman operator, denoted by $\mathcal{K}$, is an infinite-dimensional linear operator that acts on this vector space, given by \cite{koopman1931hamiltonian, budivsic2012applied, arbabi2017ergodic}:

\begin{equation}
\label{eq: sec2_3}
\mathcal{K} g \left( \boldsymbol{x} \right) = g \circ \boldsymbol{F} \left( \boldsymbol{x} \right) \Longrightarrow \mathcal{K} g \left( \boldsymbol{x}_{t} \right) = g \left( \boldsymbol{x}_{t+1} \right)
\end{equation}

where $\circ$ denotes the function composition operator. The key property of the Koopman operator exploited in this study is its linearity. It is associated with the linearity of the addition operation and follows the definition in \cref{eq: sec2_3}:

\begin{equation}
\label{eq: sec2_4}
\mathcal{K} [g_1 + g_2] \left( \boldsymbol{x} \right) = g_1 \circ \boldsymbol{F} \left( \boldsymbol{x} \right) + g_2 \circ \boldsymbol{F} \left( \boldsymbol{x} \right) = \mathcal{K} g_1 \left( \boldsymbol{x} \right) + \mathcal{K} g_2 \left( \boldsymbol{x} \right)
\end{equation}

\begin{figure}[H]
\includegraphics[width=1.0\textwidth]{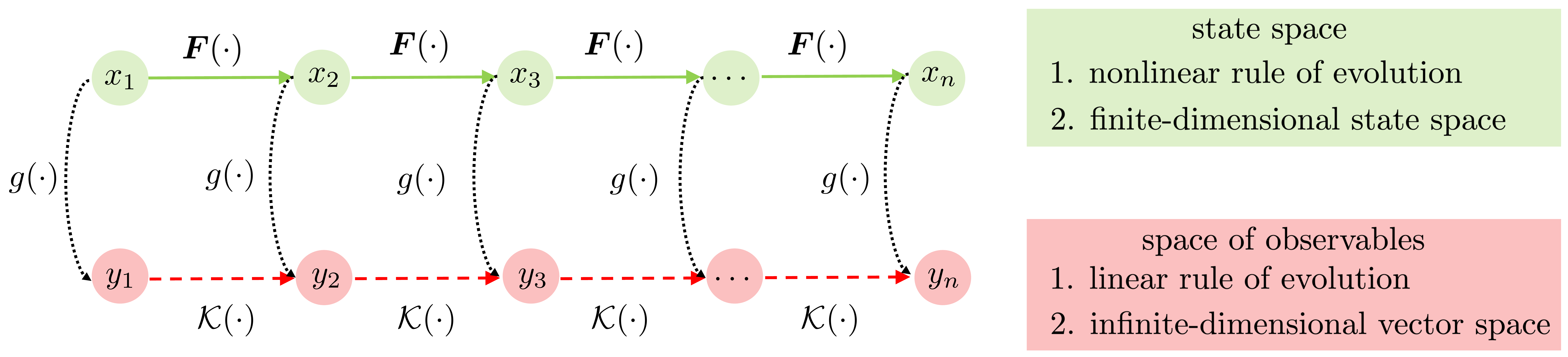}
\caption{Illustration of the dynamical evolution in the state and \textit{observable} space.} 
\label{fig: f2}
\end{figure}

Note the Koopman operator provides a linear rule of evolution that is appealing to dynamical systems analysis. However, the emerging challenge is that the space of \textit{observables} is infinite-dimensional. An schematic illustration of the Koopman operator for nonlinear dynamical systems is given in \cref{fig: f2}.

\subsection{Koopman spectral analysis}
\label{sec22}
In essence, the Koopman operator a linear operator that is unitary on a transitive invariant set. It is therefore important to study its spectral properties as this will give valuable insights into the pressure dynamics \cite{muld2012flow, zhang2014identification, mezic2013analysis}. Similar to the case of linear systems, one can view the Koopman operator as an infinite-dimensional matrix and perform eigen-decomposition of this matrix, where the computed eigenvalues and eigenmodes are physical objects in terms of explaining how the matrix acts on a vector space. Usually, the Koopman eigen-equation writes as:

\begin{equation}
\label{eq: sec2_5}
\mathcal{K}^{t} \phi_j \left( \boldsymbol{x} \right) = e^{\lambda_j{t}} \phi_j \left( \boldsymbol{x} \right) \quad \text{where} \quad j = 1, 2, \dots
\end{equation}

where $\lambda_j \in \mathbb{C}$ is the $j^{th}$ Koopman eigenvalue and $\phi_j: S \rightarrow \mathbb{C}$  denotes the paired Koopman eigenfunction, which is a complex-valued \textit{observable} of the dynamical system stated in \cref{eq: sec2_1}. For the discrete-time system, the definition is slightly different:

\begin{equation}
\label{eq: sec2_51}
\mathcal{K}^{t} \phi_j \left( \boldsymbol{x} \right) = \lambda_j \phi_j \left( \boldsymbol{x} \right) \quad \text{where} \quad j = 1, 2, \dots
\end{equation}

\subsection{Koopman mode decomposition}
\label{sec23}
The eigenvalues and eigenfunctions obtained by solving \cref{eq: sec2_51} contain spectral information about the underlying dynamical system. In particular, for systems with the Koopman spectrum consisting of only eigenvalues, the eigenfunctions of the Koopman operator form a complete basis set for the space of \textit{observables} \cite{budivsic2012applied, arbabi2017ergodic, takeishi2017learning}. Thus, with the aim of representing nonlinear dynamics using a set of eigen-measurements that evolve linearly in time, one has to span an arbitrary \textit{observable} like $g$ in terms of these Koopman eigenfunctions:

\begin{equation}
\label{eq: sec2_6}
g \left( \boldsymbol{x} \right) = \sum_{i}^{\infty} \vartheta_i \phi_i \left( \boldsymbol{x} \right)
\end{equation}

with scalars $\vartheta_i$ representing coefficients of expansion. Dynamical systems involving multiple \textit{observables}, which is more common in practice, can be similarly expanded:

\begin{equation}
\label{eq: sec2_7}
\boldsymbol{g} \left( \boldsymbol{x} \right) = \begin{bmatrix}
    g^1 \left( \boldsymbol{x} \right) \\
    \vdots \\
    g^m \left( \boldsymbol{x} \right)
\end{bmatrix} = \sum_{i}^{\infty} \boldsymbol{\vartheta}_i \phi_i \left( \boldsymbol{x} \right)
\end{equation} 

where $\boldsymbol{g}: S \rightarrow \mathbb{R}^m$ denotes a vector-valued \textit{observable} and the eigenfunctions $\phi_i$ provide a basis for Hilbert space. Given the linear Koopman expansion stated in \cref{eq: sec2_7}, one can reformulate the dynamics (\cref{eq: sec2_51}) of the measurements $\boldsymbol{g}$ as follows:

\begin{equation}
\label{eq: sec2_8}
\mathcal{K}^{t} \boldsymbol{g} \left( \boldsymbol{x} \right) = \sum_{i}^{\infty} \boldsymbol{\vartheta}_i \lambda_i \phi_i \left( \boldsymbol{x} \right)
\end{equation}

This process is known as the Koopman mode decomposition (KMD), and was introduced by Mezic in 2005 \cite{budivsic2012applied, mezic2013analysis}. It implies that when the initial state is fixed, say $\boldsymbol{x} = \boldsymbol{x}_0$, the dynamics recorded by measuring $\boldsymbol{g}$ over a vector-valued trajectory is an infinite summation (e.g. sum of sinusoids and exponentials in the context of a continuous-time system). The KMD method is particularly advantageous because the dynamics associated with each eigenfunction $\phi_i$ are determined by its corresponding eigenvalue.

\begin{remark}[1]
For conservative dynamical systems, $\boldsymbol{\vartheta}_i$ in \cref{eq: sec2_8} is a unitary Koopman mode associated with the eigen-pair $(\lambda_i, \phi_i)$. It can be computed via the direct projection of the \textit{observable} $\boldsymbol{g}$ onto the eigenfunction $\phi_i$. These modes correspond to spatial coherent structures of direct measurements (e.g. the fluctuating wind pressures) and provide unique temporal patterns, where each pattern is characterized by linear growth or decay.
\end{remark}

\subsection{Data-driven Koopman analysis}
\label{sec24}
To proceed with the linearization of nonlinear dynamics using KMD, one has to obtain the Koopman spectral properties (eigenvalues, eigenfunctions, and modes). A variety of data-driven methods have been developed to address this issue. Despite specific algorithm differences, the first step of any type of data-driven Koopman analysis is always to assemble the experimental/simulation dataset that describes the evolution of \textit{observables} \cite{taira2017modal, kutz2016dynamic}. In the context of wind pressure analysis, let \textit{observables} be the direct spatial measurements of the pressure distribution over a surface of the prism. The goal is to identify a set of Koopman tuples $(\lambda_i, \phi_i, \boldsymbol{\vartheta}_i)$ that can link the state-space parameterization $\boldsymbol{F}$ to the evolution of \textit{observables} $\mathcal{K}$ in terms of analyzing aerodynamic characteristics (See \cref{eq: sec2_3}).

\begin{figure}[H]
\centering
\includegraphics[width=1.0\textwidth]{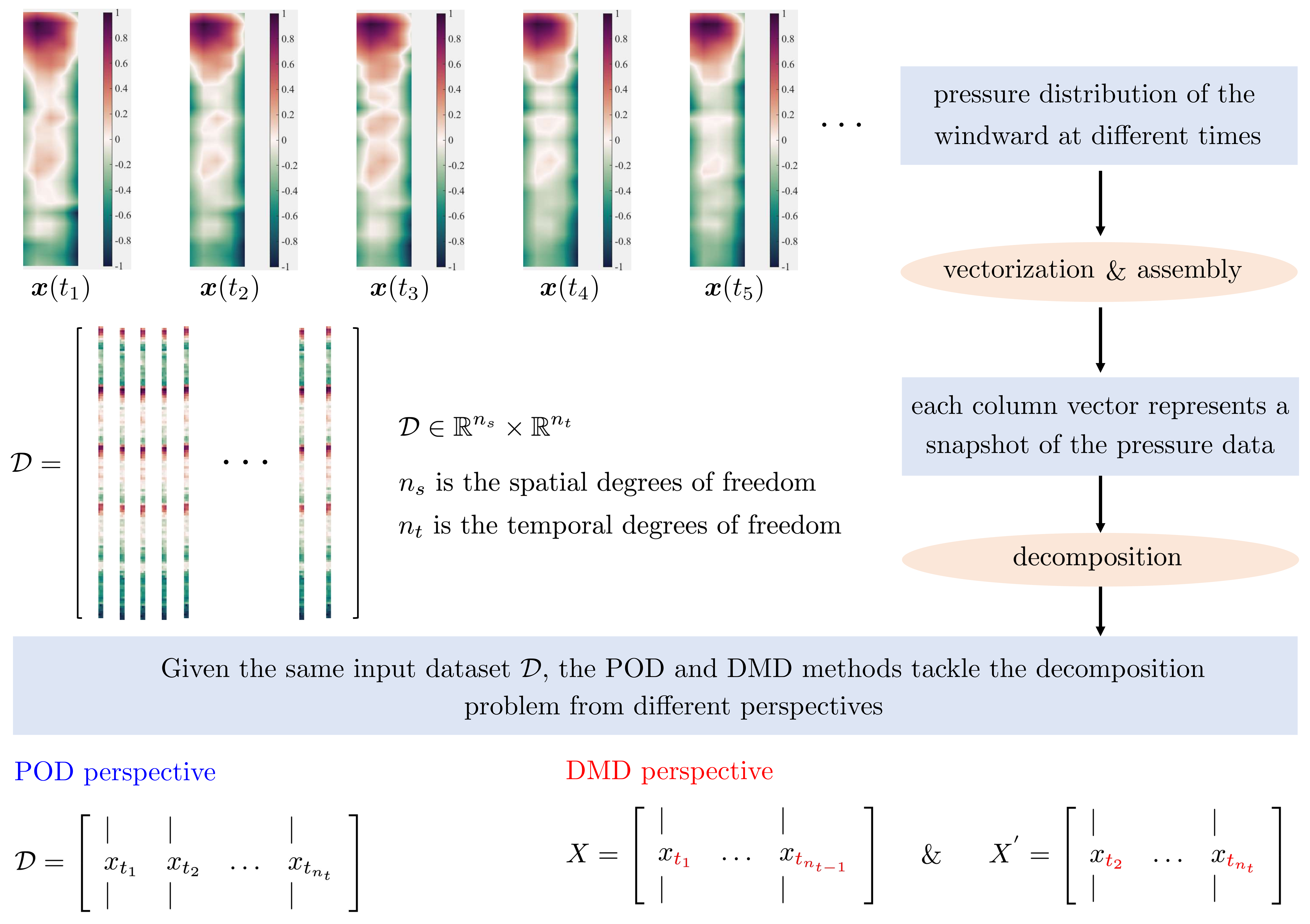}
\caption{Schematic illustration of data-driven decomposition methods.} 
\label{fig: f3}
\end{figure}

\cref{fig: f3} graphically summarizes such data collection process. First, the wind pressure is measured at a number of discretized locations. Take the windward for instance, snapshots of the pressure distribution describe the aerodynamic evolution. Second, each snapshot is vectorized into a column vector. All of the column vectors are subsequently assembled in a matrix as the input dataset $\mathcal{D}$ for decomposition. Third, similar to POD, the proposed decomposition method computes eigenmodes by solving an eigenvalue problem for a linear operator constructed from this input dataset. However, POD deals with the covariance matrix of $\mathcal{D}$, whereas the proposed method partitions $\mathcal{D}$ into a pair of time-shifted matrices ($X$ and $X^{'}$) and carries out the spectral analysis of the cross-correlation matrix that links the snapshots at time $t_{i+1}$ to the snapshots at time $t_{i}$. \cref{app1} gives the details about the POD method and the following section discusses the proposed method.

\subsection{Dynamic mode decomposition}
\label{sec25}
The dynamic mode decomposition, or DMD in short, is a data-driven regression method developed by Schmid in the fluid mechanics' community to identify spatio-temporal patterns from high-dimensional experimental data \cite{schmid2010dynamic}. In later work, Rowley et al. demonstrated that linear expansion in DMD provides a finite dimensional approximation of the KMD, where the eigenvalues and eigenfunctions obtained by the DMD algorithm converge to the Koopman eigenvalues and eigenfunctions \cite{rowley2009spectral}. Several DMD algorithms have been proposed recently \cite{chen2012variants, tu2013dynamic, le2017higher}. And the methodology adopted in this paper is the so-called exact dynamic mode decomposition (EDMD) algorithm \cite{williams2015data}. In particular, we apply EDMD to an augmented dataset, where pressure data is arranged in a Hankel-matrix type following the Takens's embedding theorem. By combining delay-embedding of fluctuating wind pressure measurements with EDMD, the dominant dynamical patterns buried in the original time series dataset have been strengthened.

The computational procedures are given in \cref{alg: EDMD} and several properties should be addressed in terms of the numerical implementation. First, each element $\boldsymbol{x}_{t_i} (i=1, \dots, n)$ in the matrix $\mathcal{D}$ is generated from a vector space $S \subset \mathbb{R}^{n_s}$, where $n_s$ denotes the spatial dimension, i.e. the number of pressure taps used in a wind tunnel test. Second, we form the Hankel matrix by stacking the elements of time-delayed coordinates, where the index $i$ indicates the embedding number. Third, the goal is to find the best-fit linear operator that connects the time-shifted matrices $X^{'} = \mathcal{A} X$, where $\mathcal{A}$ asymptotically approach the Koopman operator $\mathcal{K}$. Fourth, the singular value decomposition (SVD) is a robust algorithm that provides converged POD modes on \textit{observables} of ergodic dynamical systems. In \cref{eq: sec2_12}, $\boldsymbol{U} \in \mathbb{R}^{(n_s \times i) \times (n_s \times i)}$, $\boldsymbol{V} \in \mathbb{R}^{(j-1) \times (j-1)}$, $\boldsymbol{\Sigma} \in \mathbb{R}^{(n_s \times i) \times (j-1)}$, and $*$ denotes the conjugate transpose. $\boldsymbol{U}$, whose columns are numerical realizations of POD modes, follows $\boldsymbol{U}^{*}\boldsymbol{U}=\boldsymbol{I}$, and $\boldsymbol{V}$, whose columns are temporal behaviors, holds $\boldsymbol{V}^{*}\boldsymbol{V}=\boldsymbol{I}$ (See \cref{app1} for details). Fifth, an appropriate choice of the truncation number $r$ can be decided by means of hard-threshold computing. Sixth, the lower rank representation $\tilde{\mathcal{A}} \in \mathbb{R}^{r \times r}$ integrates the dynamical information by exploiting $X^{'}$. Seventh, the eigenvalue-eigenvector pairs for $\tilde{\mathcal{A}}$ can be obtained by solving the eigen-equation stated in \cref{eq: sec2_15}, where columns of $\boldsymbol{\Phi}$ are eigenvectors. $\boldsymbol{\Lambda}$ is a diagonal matrix with non-zero entries representing eigenvalues. The growth/decay rates and corresponding frequencies can be further determined from the eigenvalue sequence $\lambda_1, \lambda_2, \dots$ via $\sigma_i = \frac{\log \left( \mathcal{R}e \left( \lambda_i \right) \right)}{2 \pi \Delta t}$ and $\omega_i = \frac{\log \left( \mathcal{I}m \left( \lambda_i \right) \right)}{\Delta t}$, respectively. Eighth, it should be noted that the DMD modes computed by \cref{eq: sec2_16} are commonly referred to as the exact DMD modes. An alternative approach is to multiply left singular vectors by eigenvectors $\boldsymbol{\vartheta}^{'} = \boldsymbol{U}_r \boldsymbol{\Phi}$, where each column of $\boldsymbol{\vartheta}^{'}$ is so-called the projected DMD mode. The former definition is closely related to $\boldsymbol{X}$ while the emphasis of the second definition is placed on $\boldsymbol{X}^{'}$. By orthogonally projecting $\boldsymbol{\vartheta}^{'}$ to the range of $\boldsymbol{X}$, the projected DMD modes can be exactly represented via $\boldsymbol{\vartheta}$.

\begin{algorithm}[H]
\begin{algorithmic}[H]
\caption{Augmented EDMD}
\label{alg: EDMD}
\STATE{\textbf{Input:} The snapshots matrix $\mathcal{D} = [\boldsymbol{x}_{t_1}, \boldsymbol{x}_{t_2}, \dots, \boldsymbol{x}_{t_n}]$.}
\STATE{1: Build Hankel matrix:
\begin{equation}
\label{eq: sec2_10}
H = [\boldsymbol{h}_{1}, \boldsymbol{h}_{2}, \dots, \boldsymbol{h}_{j}] \triangleq \left[\begin{array}{cccc}{\boldsymbol{x}_{t_1}} & {\boldsymbol{x}_{t_2}} & {\dots} & {\boldsymbol{x}_{t_j}} \\ {\boldsymbol{x}_{t_2}} & {\boldsymbol{x}_{t_3}} & {\dots} & {\boldsymbol{x}_{t_{j+1}}} \\ {\vdots} & {\vdots} & {\ddots} & {\vdots} \\ {\boldsymbol{x}_{t_i}} & {\boldsymbol{x}_{t_{i+1}}} & {\dots} & {\boldsymbol{x}_{t_{i+j-1}}}\end{array}\right]
\end{equation}}
\STATE{2: Define a pair of time-shifted matrices $X$ and $X^{'}$:
\begin{equation}
\label{eq: sec2_11}
X = [\boldsymbol{h}_{1}, \boldsymbol{h}_{2}, \dots, \boldsymbol{h}_{j-1}] \quad \text{and} \quad X^{'} = [\boldsymbol{h}_{2}, \boldsymbol{h}_{3}, \dots, \boldsymbol{h}_{j}]
\end{equation}}
\STATE{3: Compute the singular value decomposition of $X$:
\begin{equation}
\label{eq: sec2_12}
X = \boldsymbol{U} \boldsymbol{\Sigma} \boldsymbol{V}^{*}
\end{equation}}
\STATE{4: Take the truncation:
\begin{equation}
\label{eq: sec2_13}
\boldsymbol{U}_r = \boldsymbol{U} [ :, 1:r ], \, \boldsymbol{V}_r = \boldsymbol{V} [ :, 1:r ], \, \boldsymbol{\Sigma}_r = \boldsymbol{\Sigma} [ 1:r, 1:r ]
\end{equation}}
\STATE{5: Form the lower rank representation:
\begin{equation}
\label{eq: sec2_14}
\tilde{\mathcal{A}} \triangleq \boldsymbol{U}_r^{*} X^{'} \boldsymbol{V}_r \boldsymbol{\Sigma}_r^{-1}
\end{equation}}
\STATE{6: Perform the spectral decomposition:
\begin{equation}
\label{eq: sec2_15}
[ \boldsymbol{\Phi}, \boldsymbol{\Lambda}] = eig \left( \tilde{\mathcal{A}} \right) \quad \text{where} \quad \tilde{\mathcal{A}} \boldsymbol{\Phi} = \boldsymbol{\Phi} \boldsymbol{\Lambda}
\end{equation}}
\STATE{7: Compute the DMD modes:
\begin{equation}
\label{eq: sec2_16}
\boldsymbol{\vartheta}_{k}^{r} = X^{'} \boldsymbol{V}_r \boldsymbol{\Sigma}_r^{-1} \boldsymbol{\Phi} [:, k]
\end{equation}}
\STATE{\textbf{Output:} The DMD eigen-elements.}
\end{algorithmic}
\end{algorithm}

\section{Mathematical examples}
\label{sec8}
To clearly illustrate the unique feature of using augmented dynamic mode decomposition (\cref{alg: EDMD}) to distill dynamic patterns directly from available datasets, two synthetic nonlinear systems are considered herein. Moreover, the DMD results are compared against the results obtained using the POD and independent component analysis (ICA) method \cite{carassale2011statistical, jolliffe2011principal, taira2017modal}.

\subsection{Example 1: two mixed spatio-temporal signals}
\label{sec81}
In the first example \cite{kutz2016dynamic}, the nonlinear synthesis dataset of interest can be expressed by summing the following two functions:

\begin{equation}
\label{eq: e_1}
\begin{array}{l}{f_{1}(x, t)=\operatorname{sech}(x+3) \exp (i 7 t)} \\ {f_{2}(x, t)=2 \operatorname{sech}(x) \tanh (x) \exp (i 3 t)}\end{array}
\end{equation}

\begin{figure}[H]
\centering
\includegraphics[width=1.0\textwidth]{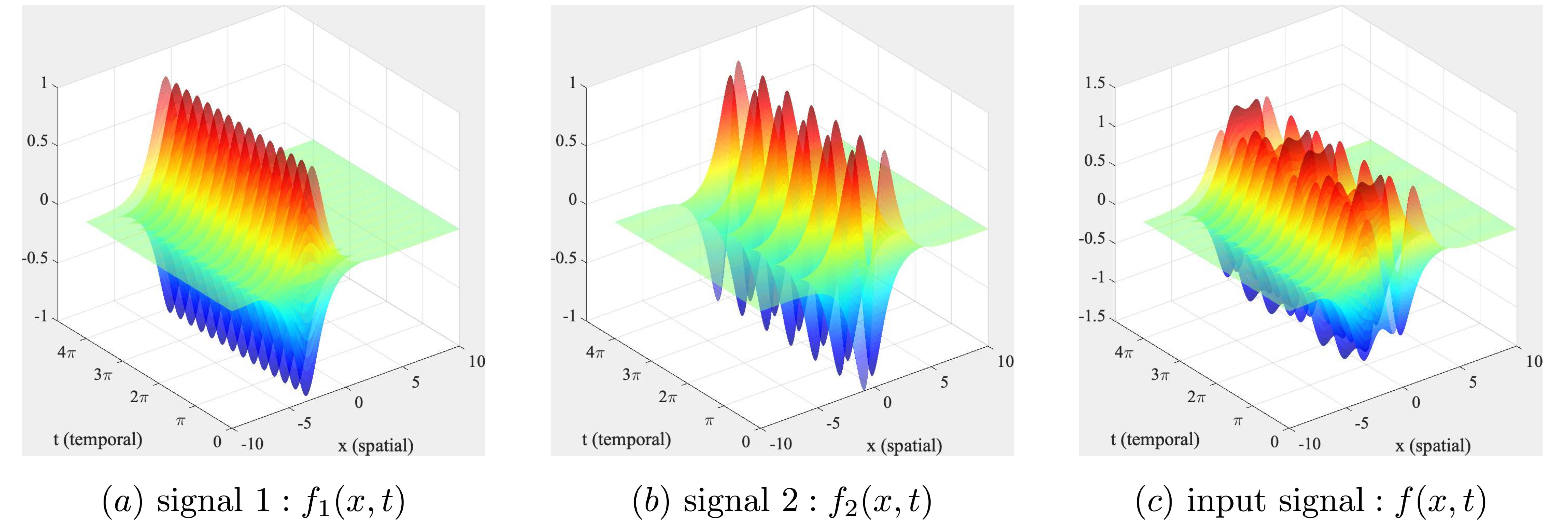}
\caption{Illustration of the components of the nonlinear dynamical system.} 
\label{fig: e1}
\end{figure}

\begin{figure}[H]
\centering
\includegraphics[width=0.9\textwidth]{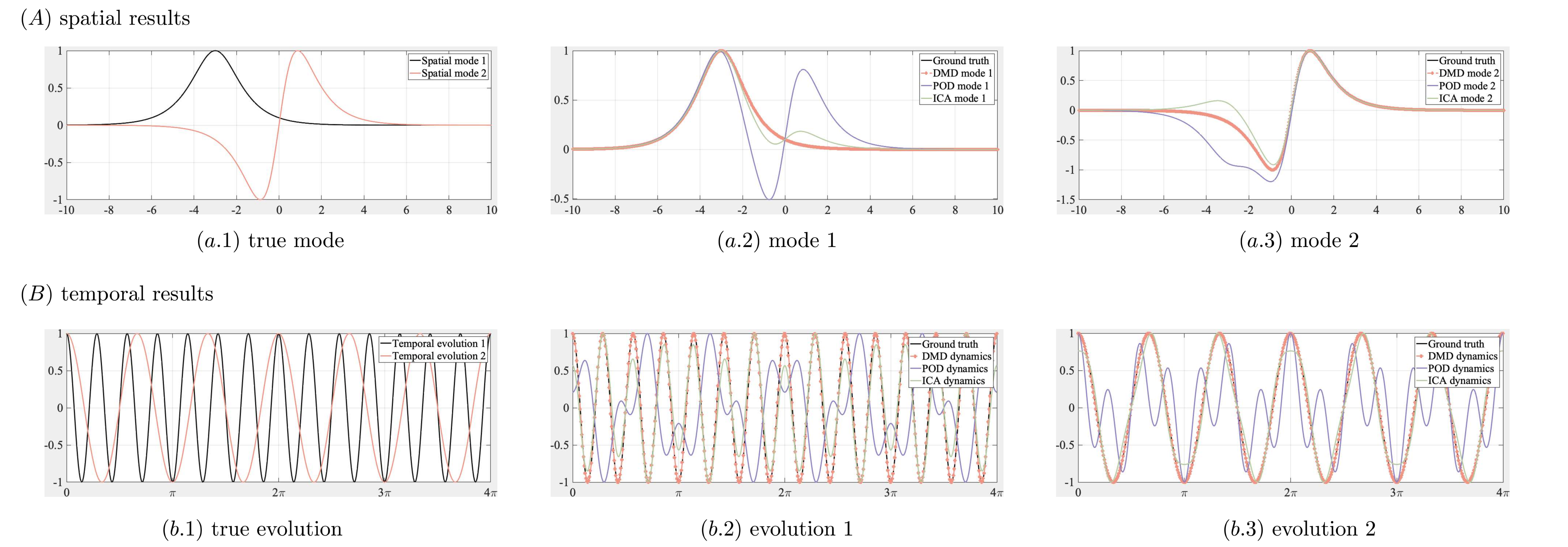}
\caption{Spatial-temporal decomposition results of 1D example.} 
\label{fig: e2}
\end{figure}

The evolving system is simulated by $f(x, t) = f_{1}(x, t) + f_{2}(x, t)$, where the two frequencies involved in the simulation are $\omega_1 = 7$ and $\omega_2 = 3$. In \cref{fig: e1}, $f_{1}(x, t)$ and $f_{2}(x, t)$ are simulated in the interval of $[0, 4\pi]$. For the purpose of comparison, the same amount of data have been fed into the DMD, POD and ICA algorithm. It should be noted that the imaginary values have been removed in the case of ICA as the whitening process included in ICA decomposition requires \cite{carassale2011statistical, taira2017modal}. The decomposed results are detailed in \cref{fig: e2}. For DMD, the truncation number specified in \cref{alg: EDMD} is initialized to $2$. The results reveal that the two dynamical patterns obtained via DMD coincide with the true evolutions and the DMD modes are almost identical to the true solution. On the other hand, POD provides a better approximation of the spatial modes than ICA while ICA excels POD in terms of learning the dynamic evolution of identified patterns.

\subsection{Example 2: mixed 2D oscillations}
\label{sec82}
In this example, a synthetic dataset with two hyperbolic oscillations is created for testing the performance of different decomposition methods. The spatial modes take the following formula:

\begin{equation}
\label{eq: e_2}
\begin{array}{l}{u_1 = \exp(-((x+10)^2+(y+10)^2)/100)} \\ {u_2 = \exp(-((x-10)^2+(y-10)^2)/100)}
\end{array}
\end{equation}

\begin{figure}[b!]
\centering
\includegraphics[width=0.9\textwidth]{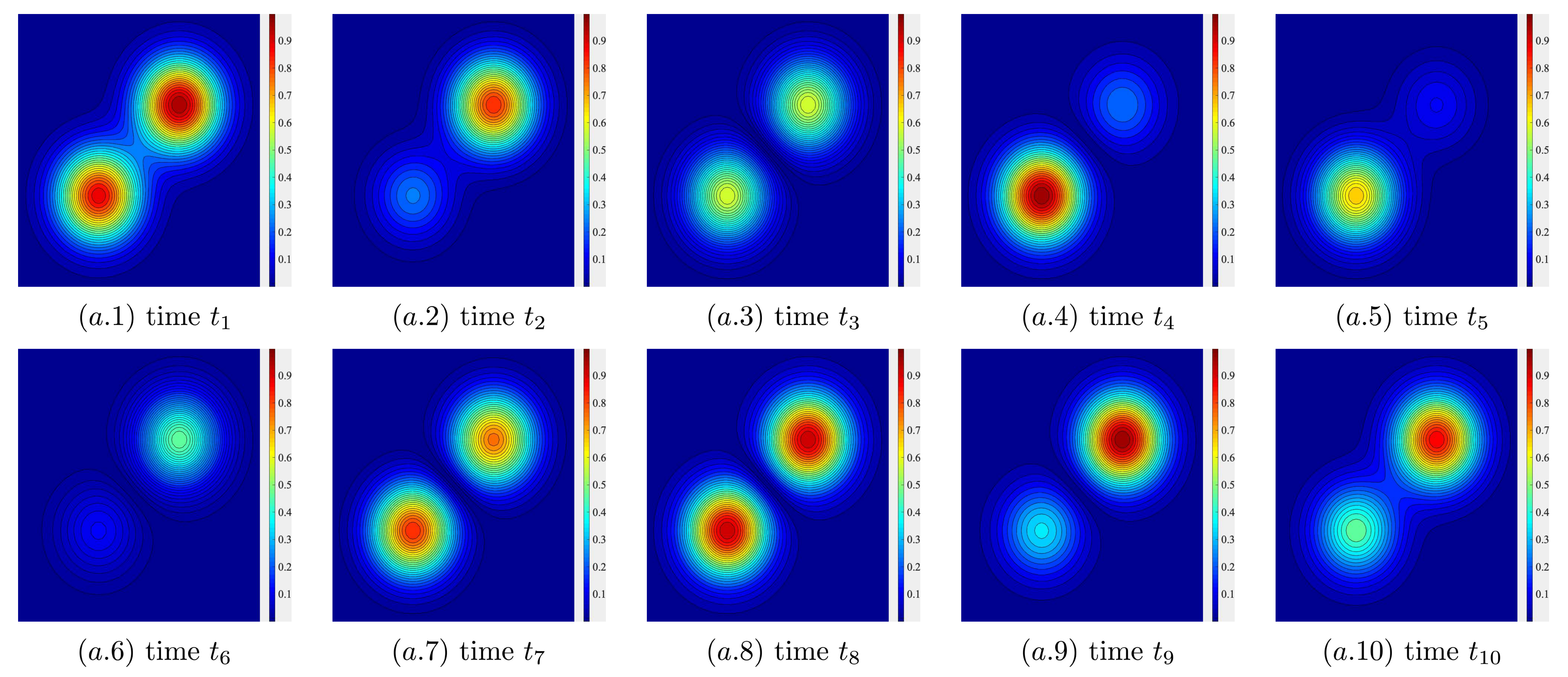}
\caption{Snapshots of the dynamical system.} 
\label{fig: e3}
\end{figure}

Specifically, the spatial domain $\Omega \subset [-30, 30] \times [-30, 30]$ with $70$ elements per axis, resulting in a resolution of $70 \times 70 = 4900$ per snapshot. The first mode $u_1$ oscillates at $7 Hz$ and the second mode $u_2$ oscillates at $3 Hz$. Note both modes are stationarily oscillate for the entire during of the simulation. The data is recorded with $\delta t = 0.01$ and a total of $1000$ snapshots have been collected for analysis. \cref{fig: e3} shows the states of the first $10$ equispaced time instances. Apparently, two modes have overlapped spatial destitutions and they are evolving at different frequencies. To accelerate the decomposition process, the method of snapshot is considered for the implementation of DMD and POD \cite{schmid2010dynamic, chen2012variants, tu2013dynamic}, that is, the input matrix is reshaped into the size of $\mathcal{D} \in \mathbb{R}^{4900 \times 1000}$. Again, DMD accurately captures the distribution of true modes compared to the POD and ICA. Beyond that, the evolution provided by the DMD algorithm matches well with the true dynamics. In terms of POD, it focuses on finding the eigenmodes that are energetically optimal regarding the reconstruction of the input data. Hence, the POD modes are very similar to the spatial average results. ICA modes are more isolated and results become ambiguous around the overlapped area.

\begin{figure}[H]
\centering
\includegraphics[width=0.91\textwidth]{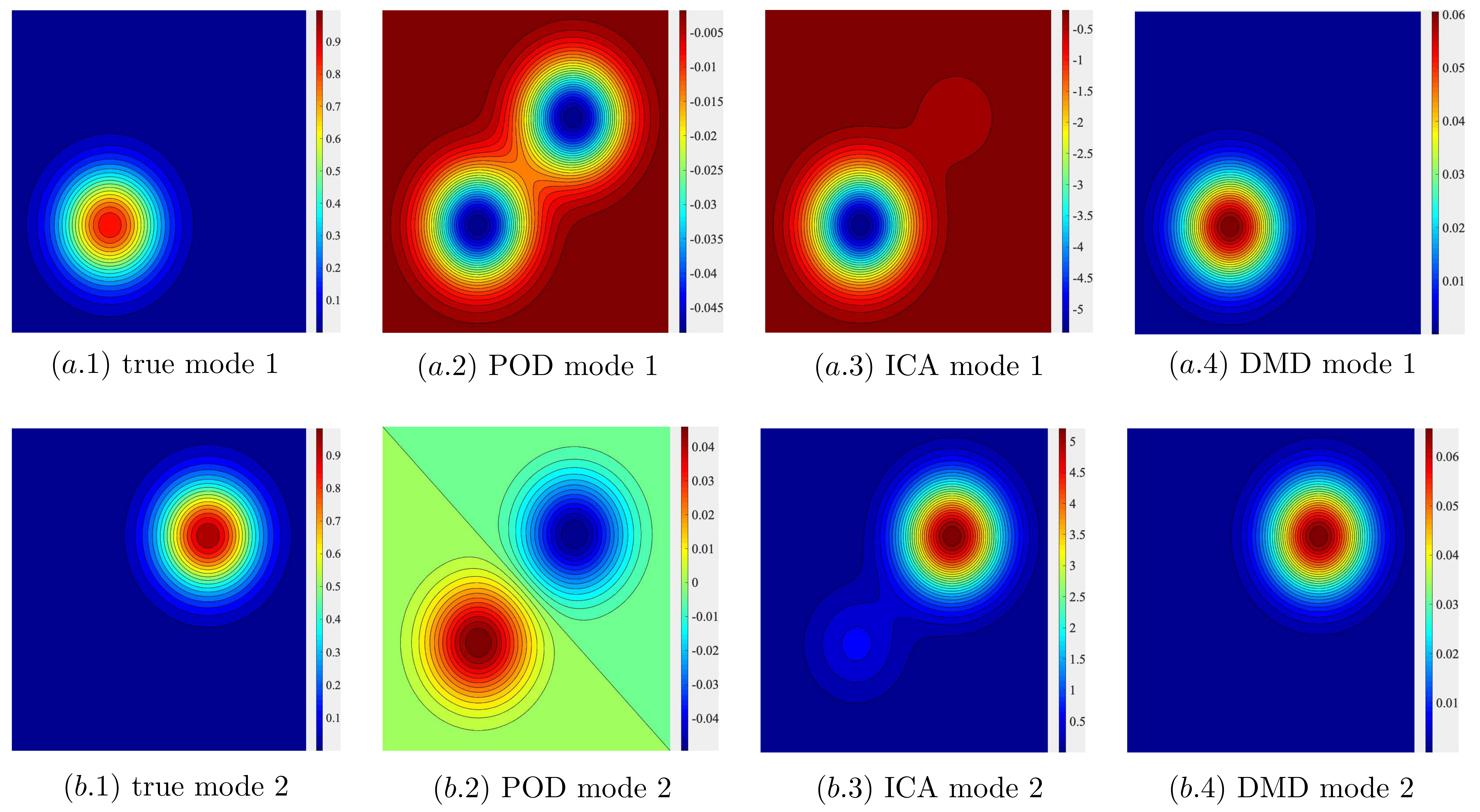}
\caption{Spatial decomposition results of 2D example.} 
\label{fig: e4}
\end{figure}

\begin{figure}[H]
\centering
\includegraphics[width=0.91\textwidth]{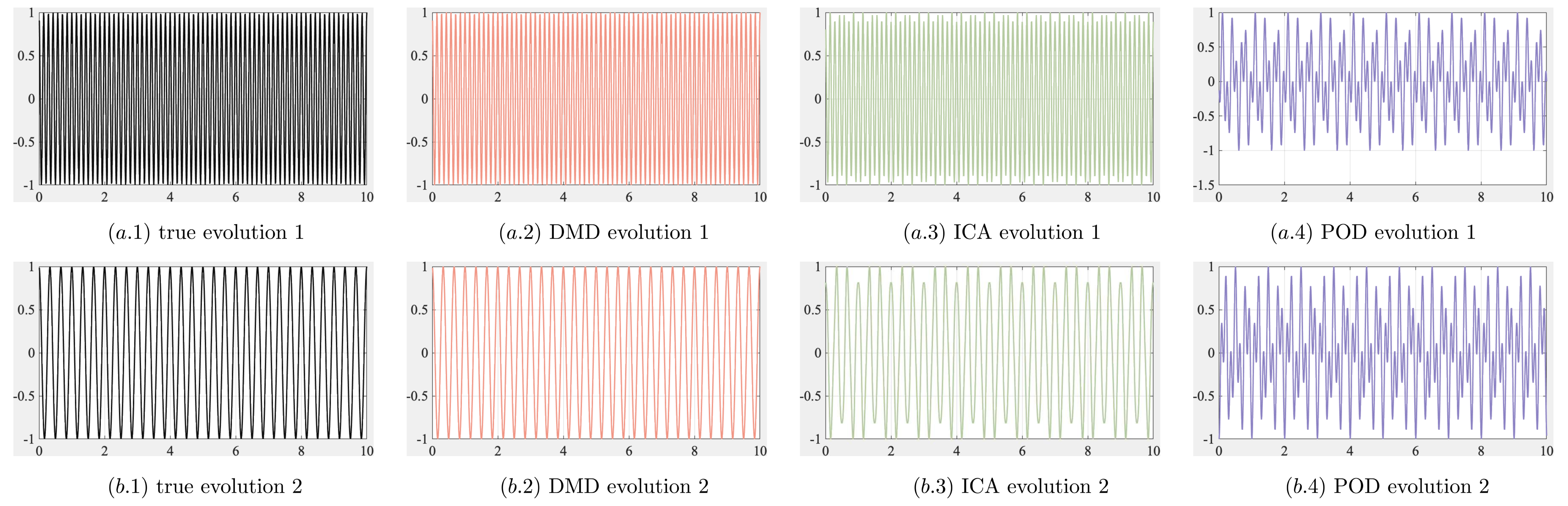}
\caption{Temporal decomposition results of 2D example.} 
\label{fig: e5}
\end{figure}

\section{Case study}
\label{sec3}

\subsection{Wind tunnel data}
\label{sec31}
In the following case study, time-series data of aerodynamic pressure fields over a finite prism ($0.5 m \times 0.1 m \times 0.1 m$) is analyzed. The experimental dataset was collected from an open-circuit wind tunnel of dimensions $14 m \times 1.2 m \times 1.0 m$ \cite{TPU}. Specifically, the approach flow mean speed profile along the height was defined by a power law exponent of $1/4$, and tests were conducted at a wind speed of $11.1438 \, m/s$. To measure the pressure, $500$ smart sensors have been installed on the scaled model. Specifically, there were $125$ sensors uniformly distributed on each surface (i.e. a  $5 \times 25$ distribution pattern in terms of the horizontal and vertical axis, respectively). All taps were synchronously recorded at a sampling frequency of $1000 \, Hz$ for a sample period of $32.768 \, s$. The collected pressure data was transformed into a dimensionless number, that is, the pressure coefficient , and matrices $X$, $X^{'}$ were accordingly assembled using \cref{eq: sec2_11}. For more information about the experimental data, please refer to \href{http://wind.arch.t-kougei.ac.jp/system/eng/contents/code/tpu}{TPU database}.

\subsection{Case study outline}
\label{sec32}
To demonstrate the unique features of using the Koopman operator to describe a random pressure field, we carried out a comparative study. Three types of analyses have been performed.

\textbf{\textsl{Convergence Study}} The objective of the convergence study is to determine how many snapshots are required to ensure the decomposition results are not affected by changing the number of snapshots \cite{muld2012flow}. In the case study, the input dataset $\mathcal{D}$ consists of $32768$ pressure states. Thus, a reasonable snapshot number $N_{snap}$ is crucial since a small $N_{snap}$ (e.g. $N_{snap} = 30$) cannot achieve converged results while a large $N_{snap}$ (e.g. $N_{snap} = 30000$) would call for an additional computation. Two evaluation metrics are defined to measure the convergence behavior (See \cref{app2}).

\textbf{\textsl{Sufficiency Analysis}} The sufficiency analysis of POD centers on constructing a low-rank representation of the fluctuating wind pressures. For the proposed augmented EDMD (See \cref{alg: EDMD}), one has to select an embedding number in such a way that the input dataset contains sufficient information for the decomposition analysis \cite{le2017higher}.

\textbf{\textsl{Modal Analysis}} The goal of the modal analysis in the context of analyzing aerodynamic pressure fields is to determine the relationship between the aerodynamic loading and the identified eigenmodes. We explained DMD results from both spatial and temporal perspectives, and connected DMD modes with the turbulent flow environment around the prism.

In this study, pressure data of the wind direction $\alpha = 0^{\circ}$ was used. The complete dataset includes wind-induced pressure coefficients on four surface of the prism. For the sake of brevity, decomposition results of the windward surface have been summarized in this paper. We report the complete results in the supplementary material, which is available at \url{https://github.com/Xihaier/DMD-POD-aerodynamic-pressure-analysis}.

\section{Part 1: convergence study results}
\label{sec4}
\subsection{POD Results}
\label{sec41}
\begin{figure}[H]
\centering
\includegraphics[width=1.0\textwidth]{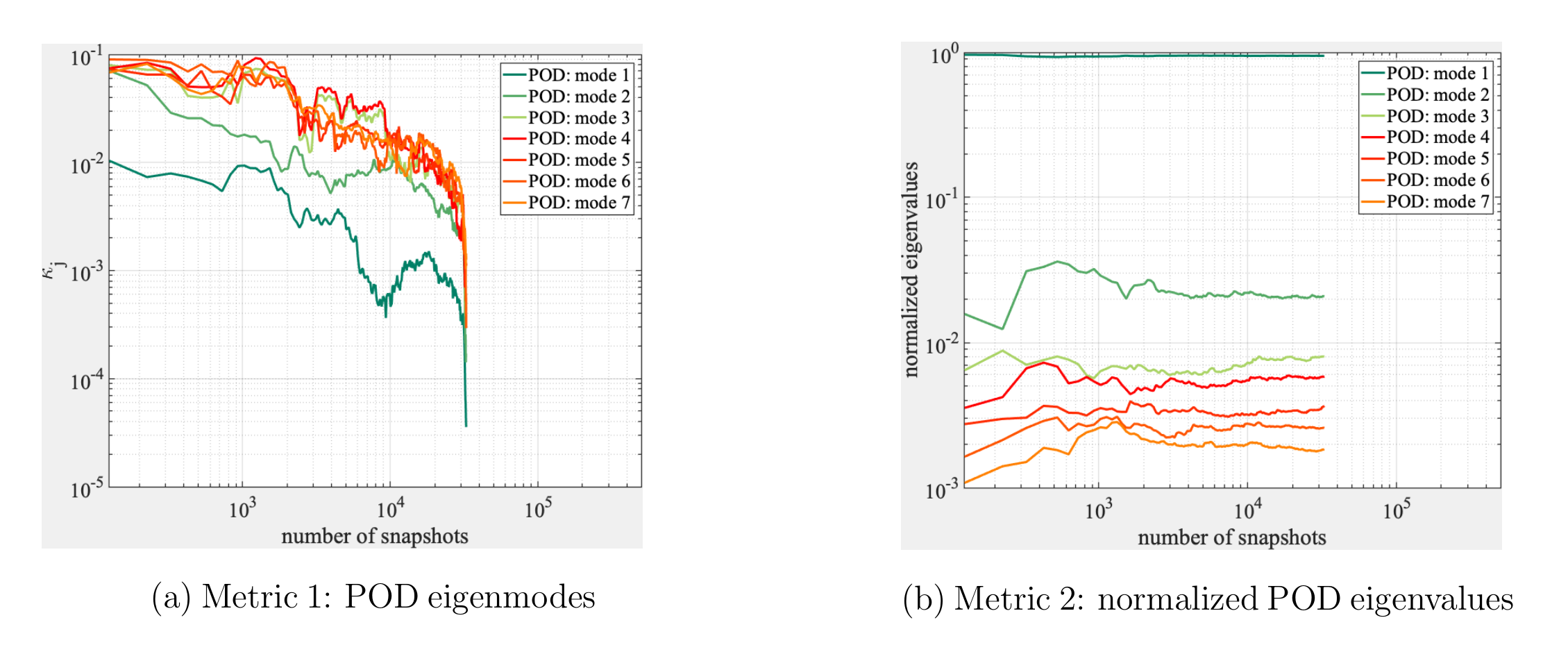}
\caption{POD convergence results of windward scenario.} 
\label{fig: f4}
\end{figure}

\cref{fig: f4} shows the convergence behavior of POD analysis. The first index based on the computed eigenmodes points out the first three fundamental POD modes $\phi_1^{POD}, \phi_2^{POD}, \phi_3^{POD}$ exhibit consistent convergence properties. The evaluation metric $\boldsymbol{\kappa}$ decreases dramatically within the time span $[ 0, 1000 \Delta t ]$. On the contrary, higher POD modes $\phi_j^{POD} \left( j = 4, \dots, 7 \right)$ not only converge at a slightly slower pace but also show more random fluctuations during the convergence process. This reveals the fact that incoherent noise tend to dominate and even likely appear as higher order POD modes \cite{aubry1988dynamics, berkooz1993proper}. Thus, the POD method can be employed as a denoising technique for preprocessing received signals that are strongly contaminated with noise \cite{solari2007proper, taira2017modal}. 

On the other hand, the second index that is defined as a function of the computed eigenvalues confirms the presence of such noise effects. The first few normalized POD eigenvalues converge at a faster speed with less local perturbation than those held by higher POD modes. Meanwhile, it is known that the normalized eigenvalues $\lambda_j^{\dagger}$ represent the mean energy distribution of the aerodynamic pressure data. Shown in \cref{fig: f4}, some POD modes have similar energy distribution as their corresponding converged eigenvalues are almost identical. From an algorithmic perspective, it is due to the data projection procedure involved in the POD analysis. This fact implies the existence of second-order statistics, i.e., correlations between projected modes \cite{jolliffe2011principal}. 

\subsection{DMD Results}
\label{sec42}
\begin{figure}[H]
\centering
\includegraphics[width=1.0\textwidth]{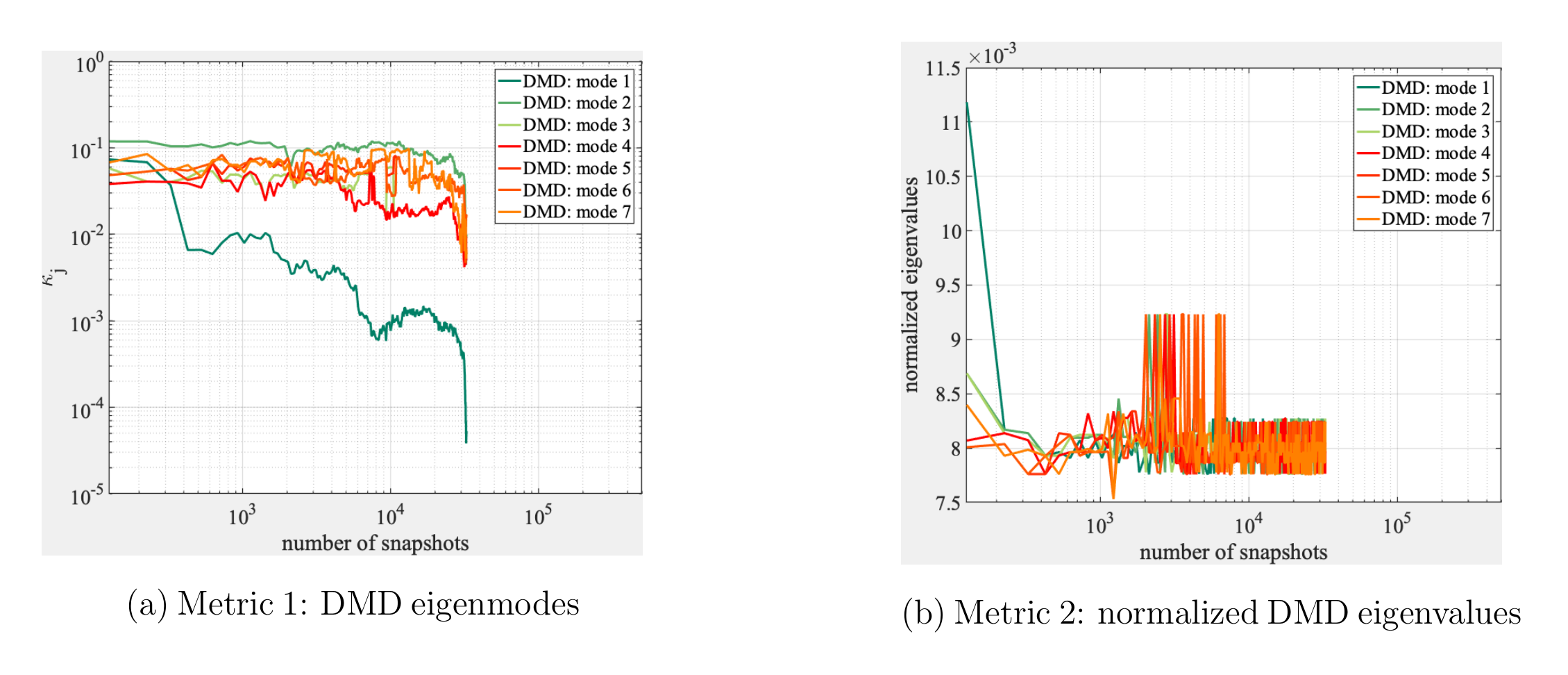}
\caption{DMD convergence results of windward scenario.} 
\label{fig: f5}
\end{figure}

\cref{fig: f5} shows the computed $\kappa_j^{DMD}$ and $\lambda_j^{\dagger}$. According to the results, neither DMD modes nor the normalized DMD eigenvalues converge during the analysis. This phenomenon is caused by the DMD algorithm, where dynamics are integrated via a pair of time-shifted matrices $X$ and $X^{'}$ \cite{schmid2010dynamic, chen2012variants}. When the size of $X$ is relatively small, DMD tends to produce a vastly expanded frequency band, where the amplitude spectra covers more isolated frequencies. As the size of $X$ increases, the observed oscillation frequencies of identified DMD modes cluster themselves in a narrowband. Thus, the observed frequencies may overlap with each other, thereby obscuring the convergent process.

\begin{figure}[H]
\centering
\includegraphics[width=1.0\textwidth]{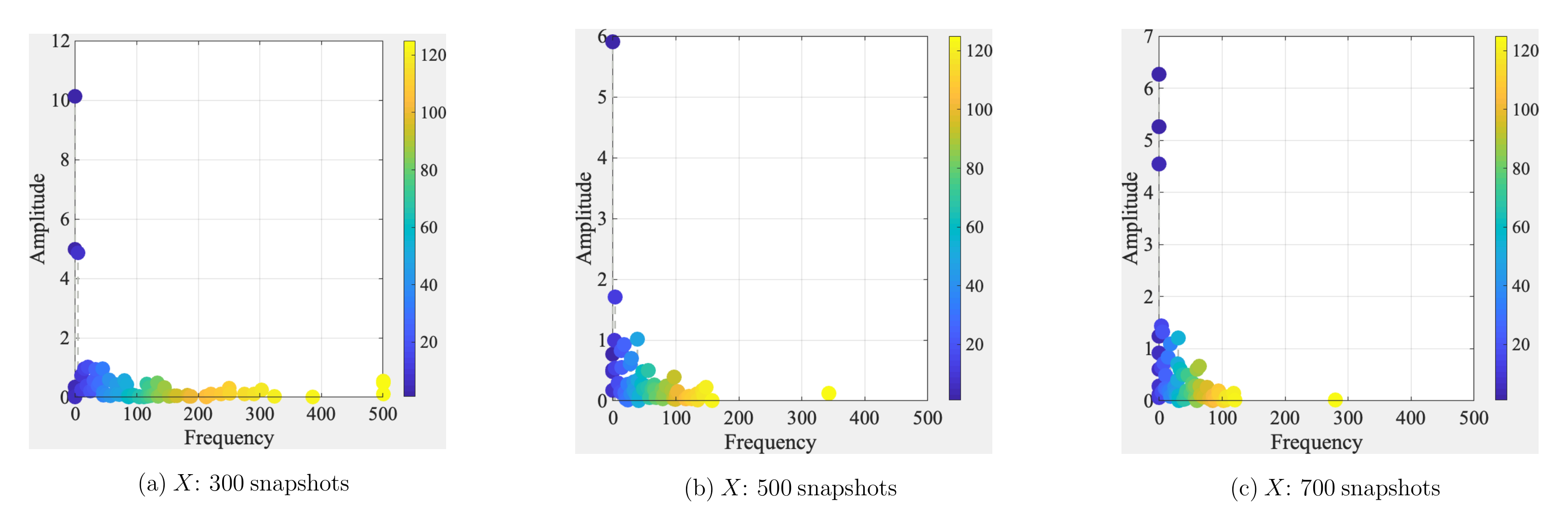}
\caption{Convergence of the DMD frequency band.} 
\label{fig: f6}
\end{figure}

\textbf{\textsl{Convergence of the frequency band}} To demonstrate the aforementioned shrinkage property, we computed the DMD energy spectrum using a different number of snapshots. In \cref{fig: f6}, it can be seen that the frequency band shrinks rapidly when $N_{snap}: 300 \rightarrow 500 \rightarrow 700$. DMD modes are sparsely scattered throughout the interval $[0, 500]$ when $N_{snap} = 300$ while the majority of resolved DMD modes are narrowed within a smaller interval $[0, 100]$ when $N_{snap} = 700$. Specifically, some eigenmodes identified by the DMD method may highly resemble each other when $N_{snap}$ is large enough. These modes share a global pattern in terms of explaining the aerodynamic pressure, but more importantly, they provide different interpretations at a local scale, which are closely related to the small scale turbulence effects. Therefore, the DMD method have achieved the convergence of the frequency band rather than the converged DMD eigenvalues and eigenmodes. 

\begin{figure}[H]
\centering
\includegraphics[width=1.0\textwidth]{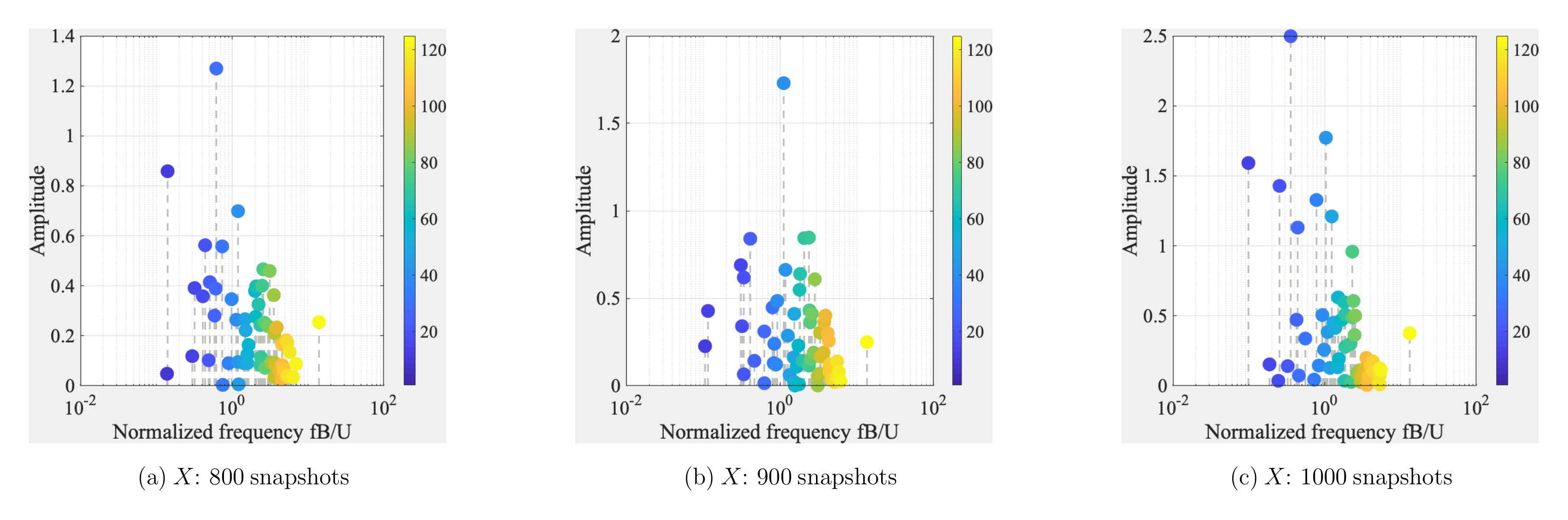}
\caption{Illustration of the dynamics in the converged frequency band.} 
\label{fig: f7}
\end{figure}

\textbf{\textsl{High/low frequency dynamics}} To investigate the dynamics in the converged narrowband, we examined the DMD energy spectrum with $N_{snap} = 800, 900, 1000$. For the illustration purpose, frequencies have been normalized by the geometric parameter $B$ and hourly average wind speed $U$. The computed results confirm the convergence of the DMD frequency band. In \cref{fig: f7}, the low-frequency region (blue dots) exhibits more distribution variation while the high-frequency region (yellow dots) is more unified and stable in against the $N_{snap}$ change. This observation agrees with the wind environment around the prism, where large eddies of lower frequencies are more energetically unstable \cite{leonard1975energy, sirovich1987turbulence, holmes2012turbulence}. In fact, each dot has a corresponding spatial pattern, which is affected by a multitude of eddies of different sizes and strengths. Due to the richness of turbulence dynamics, the decomposition results vary from case to case. This again explains the dependence of the DMD method on the $N_{snap}$ value. 

\section{Part 2: sufficiency analysis results}
\label{sec5}
\subsection{Sufficiency of the POD eigenvalues}
\label{sec51}
\begin{figure}[H]
\centering
\includegraphics[width=1.0\textwidth]{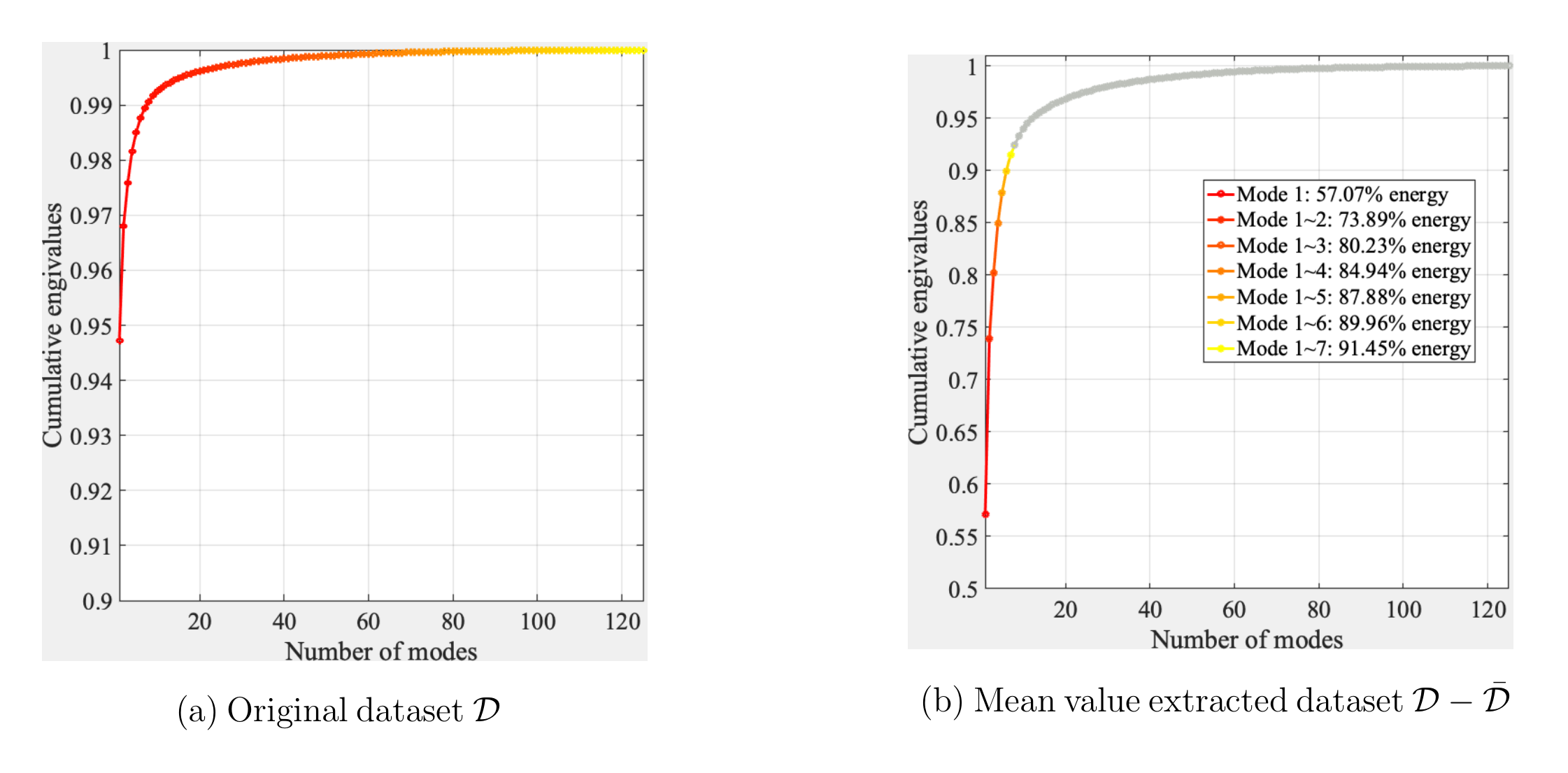}
\caption{Cumulative eigenvalues versus the number of POD modes.} 
\label{fig: f8}
\end{figure}

According to the convergence study results, the use of $1000$ successive snapshots is able to produce well-converged POD modes (See \cref{fig: f4}). Thus, we performed the POD analysis using $1000$ snapshots. \cref{fig: f8} shows the sufficiency analysis results. The ratio of the cumulative sum of the POD eigenvalues is presented, where each eigenvalue is normalized by the total energy summation $\sum_{i=1}^{125} \lambda_i$. When the original dataset $\mathcal{D}$ is used, the first eigenvalue ranked by the magnitude captures more than $90$ percent of the energy buried in $\mathcal{D}$. In many other studies, it was found that the first resolved POD mode is highly closed to the mean pressure distribution at different locations \cite{kareem1984pressure, tamura1999proper, chen2005proper, carassale2011statistical}. To investigate the aerodynamic fluctuation effects, we further extracted the mean pressure and applied POD to the remaining data $\mathcal{D} - \bar{\mathcal{D}}$. In the windward scenario, $7$ POD modes are sufficient to reconstruct a fluctuation dataset that contains $90$ percent of the energy of $\mathcal{D}$. Following a similar exponential decay pattern, a need for more POD modes is required in the other three scenarios (See the supplementary material for more details). The difference reveals the fact that the intrinsic complexity of the aerodynamic pressure field on the prism surface is strongly related to the incoming turbulence, which comes from a wide range of scales. 

\subsection{Sufficiency of the delay-embedding and DMD eigenvalues}
\label{sec52}
\begin{figure}[H]
\centering
\includegraphics[width=0.8\textwidth]{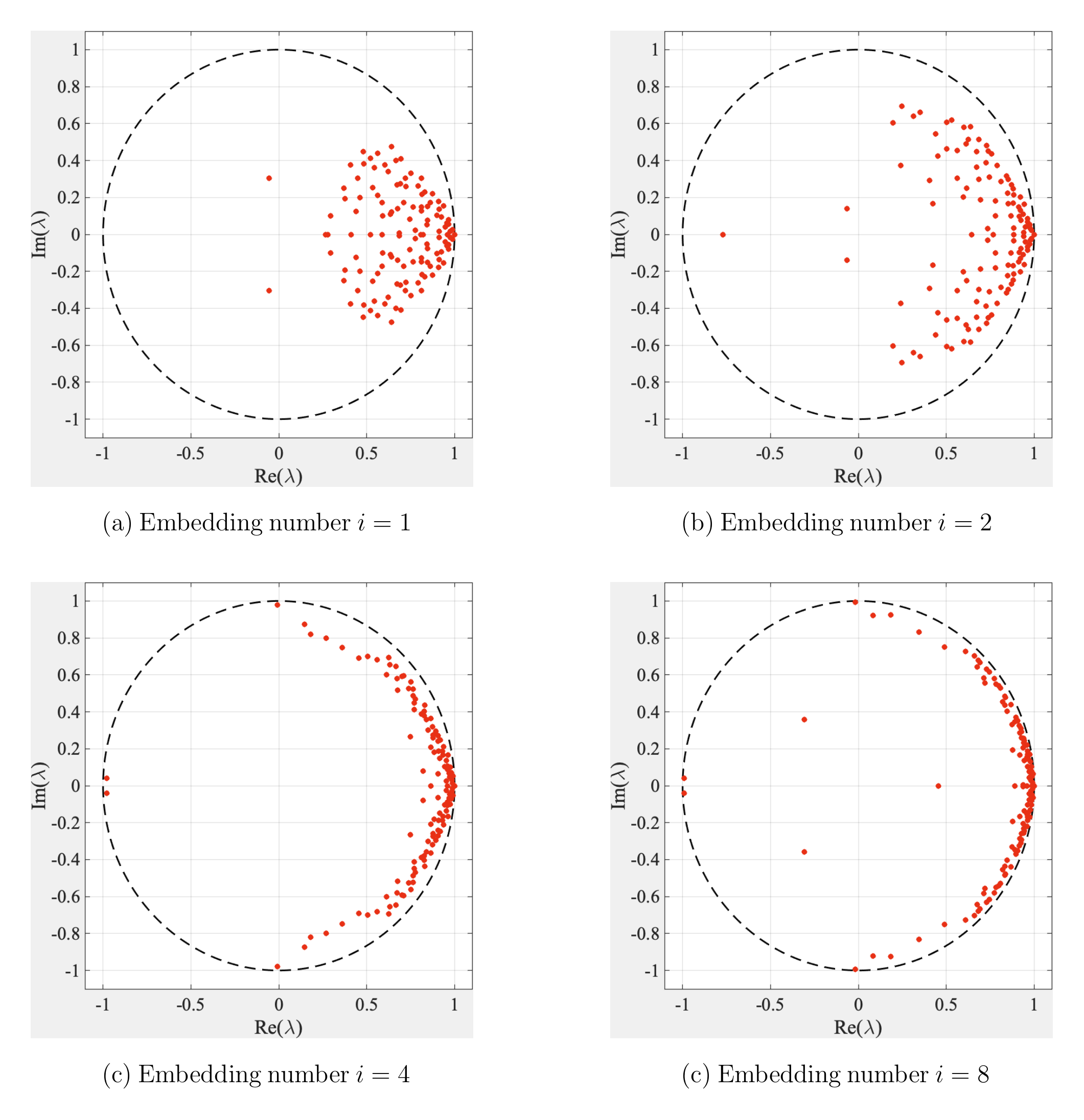}
\caption{Scatterplot of the DMD eigenvalues $\lambda_1, \dots, \lambda_{125}$.} 
\label{fig: f9}
\end{figure}

To check whether the input dataset contains sufficient information for the DMD analysis,
we compared the decomposition results of different embedding number. \cref{fig: f9} shows the distribution of the computed DMD eigenvalues $\lambda_1, \dots, \lambda_{125}$. In practice, eigenvalues are found to be unstable if their complex modulus is greater than $1$ (i.e., are located outside the unit circle) and eigenvalues are deemed as stable or neutrally stable if $\sqrt{\mathcal{R}e \left( \lambda_i \right)^2 +  \mathcal{I}m \left( \lambda_i \right)^2} \leqslant 1$ (i.e., are located inside or on the unit circle) \cite{williams2015data, arbabi2017ergodic, tu2013dynamic, kutz2016dynamic}. According to the results, the DMD eigenvalues are scattered inside the circle when the delaying coordinates are not considered $i=1$. Such distribution indicates the decomposition algorithm needs more information in terms of separating different types of dynamics because ideally the eigenvalues are expected to be located at the circular perimeter. To address this issue, the input dataset is augmented by the Takens embedding theorem \cite{takens1981detecting}. When the embedding number $i: 1 \rightarrow 2 \rightarrow 4 \rightarrow 8$, it can be seen that the eigenvalues $\lambda_1, \dots, \lambda_{125}$ are gradually pushed to the circle boundary.

\begin{figure}[H]
\centering
\includegraphics[width=0.8\textwidth]{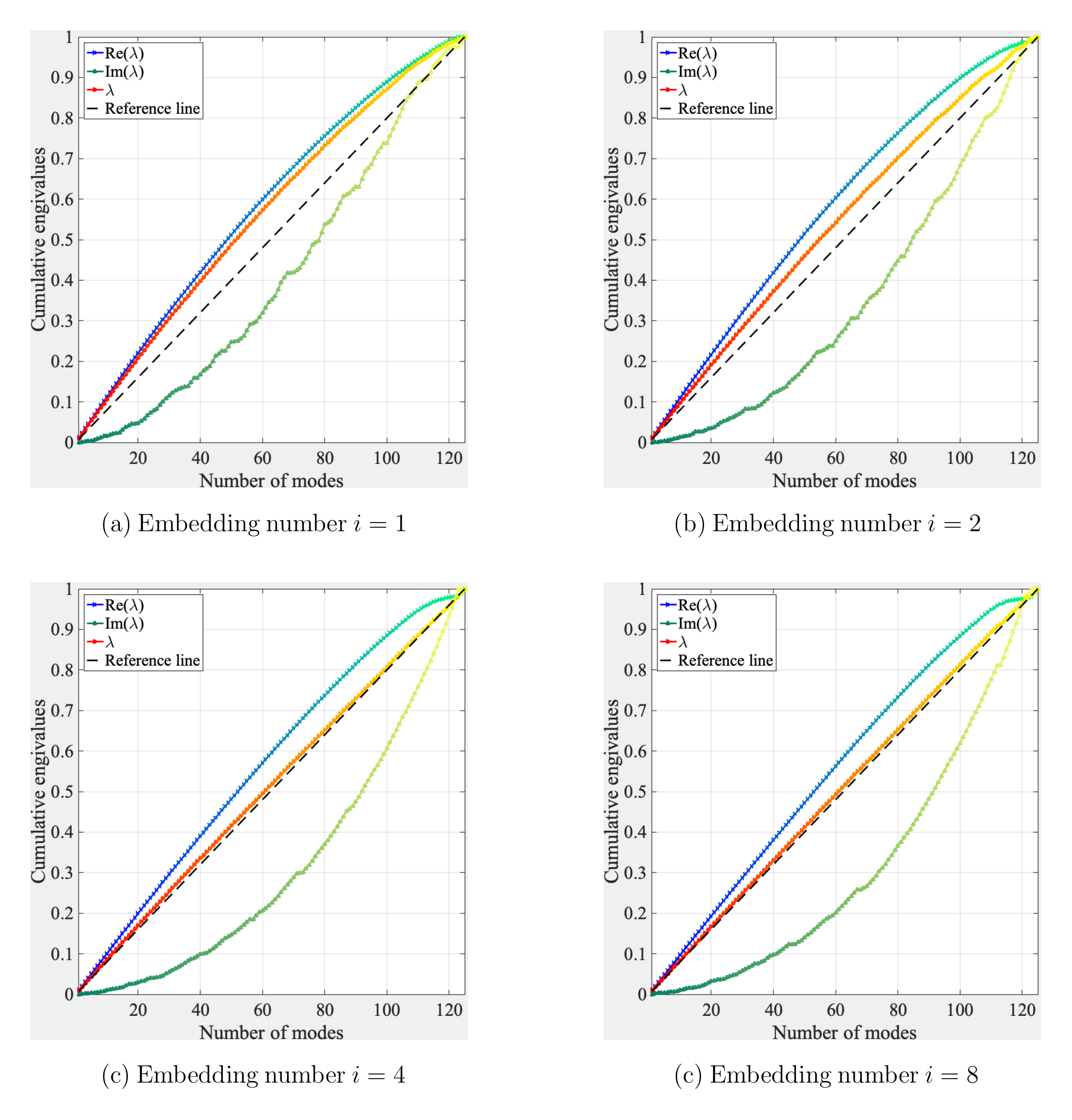}
\caption{Cumulative eigenvalues versus the number of DMD modes. The straight reference line is a linear function whose independent variable is the number of DMD modes, the dependent variable is the normalized cumulative eigenvalues, and the intercept is $0$.} 
\label{fig: f10}
\end{figure}

Similar to the sufficiency analysis of the POD eigenvalues, we ranked the complex-valued sequence $\lambda_1^{DMD}, \dots, \lambda_{125}^{DMD}$ by the absolute magnitude of the real part in descending order and plotted the cumulative sum of DMD eigenvalues. The results are summarized in \cref{fig: f10} and several properties should be noted. First, the red line representing the complex modulus summation pattern gradually converges to diagonal reference line when the embedding number increases. The reference line is a simple linear function with zero intercepts and the slope is determined by the truncation number $i$, i.e. $y = 1/125 x$. Unlike POD results, rapid growth of the cumulative eigenvalues is undetected in ordered DMD results $\mathcal{R}e \left( \lambda_1^{DMD} \right) > \mathcal{R}e \left( \lambda_2^{DMD} \right) > \dots$ (See the blue lines in \cref{fig: f10}). This is due to the fact that the nonlinear aerodynamics is spanned by an infinite Koopman modes. It is, therefore, energies of the computed Koopman modes asymptotically converge to a uniform distribution and the cumulative summation becomes linear \cite{arbabi2017ergodic, mezic2013analysis}. Second, both real and imaginary lines appear to be smoother when $i$ becomes large, affirming the enrichment effects brought by the delay coordinates. Third, the green line that covers the imaginary information of DMD eigenvalues is a downward convex function. Subsequently, the gradients of this function correspond to the phase information. It can be seen that the identified DMD modes with larger energy content tend to evolve at a slower pace than the low energy modes. This is similar to the turbulence where large eddies break into small eddies, which grow with a relatively higher frequency \cite{tennekes1972first, leonard1975energy}.

\section{Part 3: modal analysis results}
\label{sec6}
\subsection{POD modes}
\label{sec61}
\begin{figure}[H]
\centering
\includegraphics[width=1.0\textwidth]{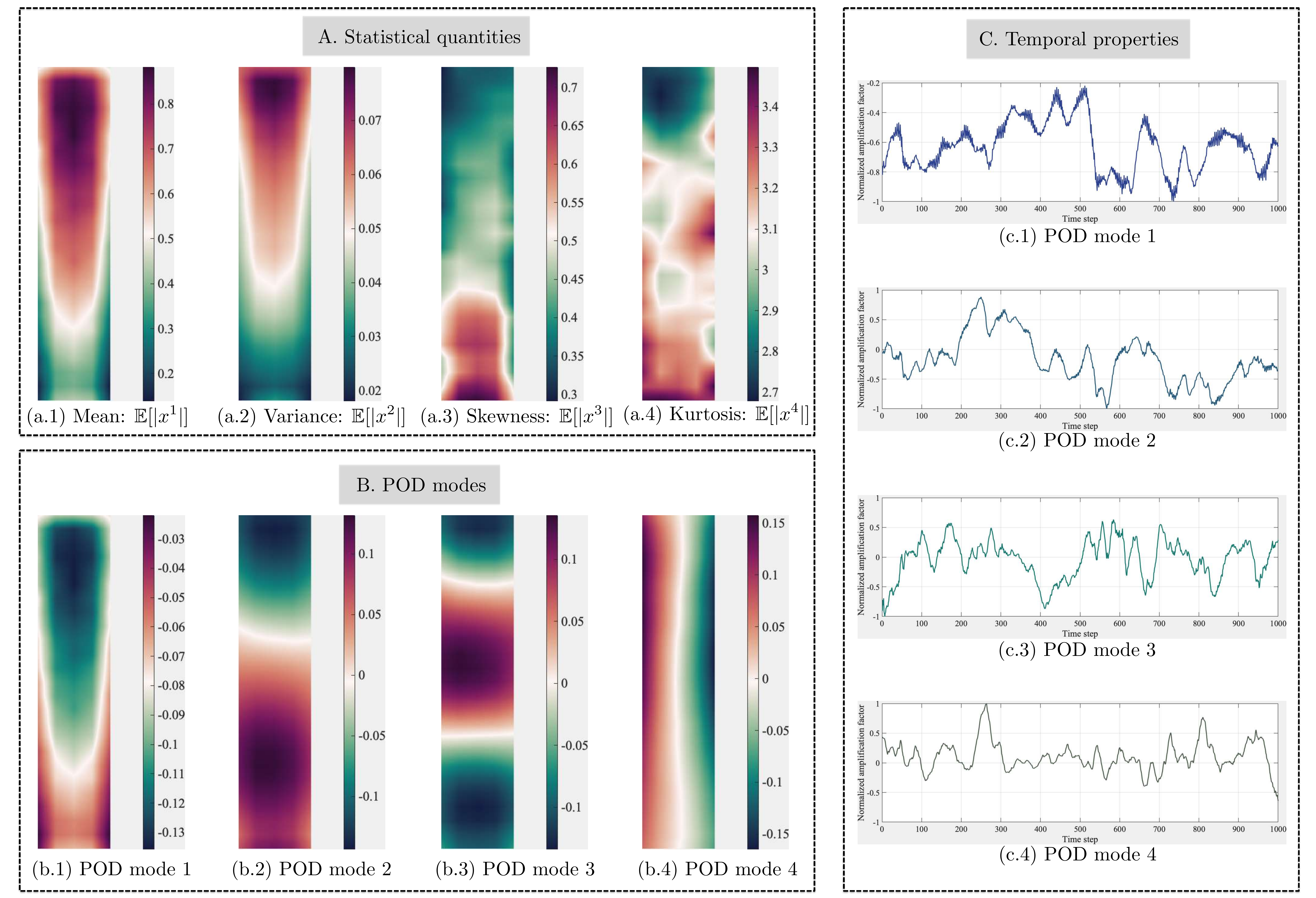}
\caption{Windward scenario: the decomposed spatial patterns and their time coefficients.} 
\label{fig: 11}
\end{figure}

\textbf{\textsl{Connections with statistical moments}} POD analysis by definition relies on the second-order statistical information \cite{jolliffe2011principal}. To show the connection between POD modes and statistical moments, we summarized the mean, variance, skewness, and kurtosis of the fluctuating wind pressures in \cref{fig: 11}. (A). It can be seen that the the first POD mode is very similar to the first-order statistical moment and the second-order statistical moment may be generally accessed via the combination of a set of POD modes. However, variability properties included in the skewness and kurtosis cannot be explained using POD modes. From a statistical learning perspective, the POD algorithm can capture the location and variability of the data but have difficulty in describing the symmetry and flatness properties.

\textbf{\textsl{Spatial interpretation of POD modes}} Singular value decomposition (SVD) is adopted to find an orthogonal set of eigenfunctions and the computed POD modes are shown in \cref{fig: 11}. (B). In common practice, POD modes are assumed to be related to the energy distribution of random pressure field \cite{lee1975effect, kareem1984pressure, tamura1999proper, solari2007proper}. For instance, $\phi_1, \phi_2$, and $\phi_3$ represent the pressure distributions that result in the along-wind overturning moment while $\phi_4$ exhibits a pressure distribution pattern that is related to torsional loads. Moreover, higher-order POD modes are more associated with small scale features since eigenmodes are arranged in decreasing order of eigenvalues. For example, peak values of $\phi_4$ are found around the prism edges. Similar results have been observed in the leeward and two sidewards where min-max values are around corners, suggesting the locates where large pressure fluctuations take place (See the supplementary material for more details). 

\textbf{\textsl{Temporal interpretation of POD modes}} The time coefficients associated with the computed POD modes are presented in \cref{fig: 11}. (C). At first glance, the evolution process looks like stochastic in nature rather than a well-ordered process. For instance, $a_2 \left( t \right), a_3 \left( t \right)$, and $a_4 \left( t \right)$ are changing abruptly in time, precluding identification of qualitative description of aerodynamic characteristics. This is caused by the zero-time-lag covariance matrix used in the POD analysis. Because such covariance matrix does not contain any evolution information, the decomposition results in assigning an eigenmode with multiple frequencies in such a way that the $L_2$ norm error in the POD formulation is minimized \cite{zhang2014identification, taira2017modal, towne2018spectral}. Consequently, the computation process precludes the prediction of the next state in time.

\subsection{DMD modes}
\label{sec62}
\textbf{\textsl{DMD setup}} To compute DMD modes, we defined parameters involved in the augmented EDMD algorithm utilizing the convergence and sufficiency analysis results presented earlier. First, $730$ snapshots have been used, that is, the input dataset $\mathcal{D} \in \mathbb{R}^{125 \times 730}$. Next, the input dataset was embedded into a higher dimension to ensure that most eigenvalues are driven to a neutrally stable state. In particular, the embedding number $i = 30$ and the embedded dataset $\mathcal{D} \in \mathbb{R}^{3750 \times 701}$. According to \cref{alg: EDMD}, the augmented decomposition algorithm would divide $\mathcal{D}$ into a pair of time-shifted matrices $\boldsymbol{X}, \boldsymbol{X}^{'} \in \mathbb{R}^{3750 \times 700}$. We fixed the truncation number $r = 125$. As a result, a low-rank representation consisting of $125$ DMD modes has been extracted.

\begin{figure}[H]
\centering
\includegraphics[width=1.0\textwidth]{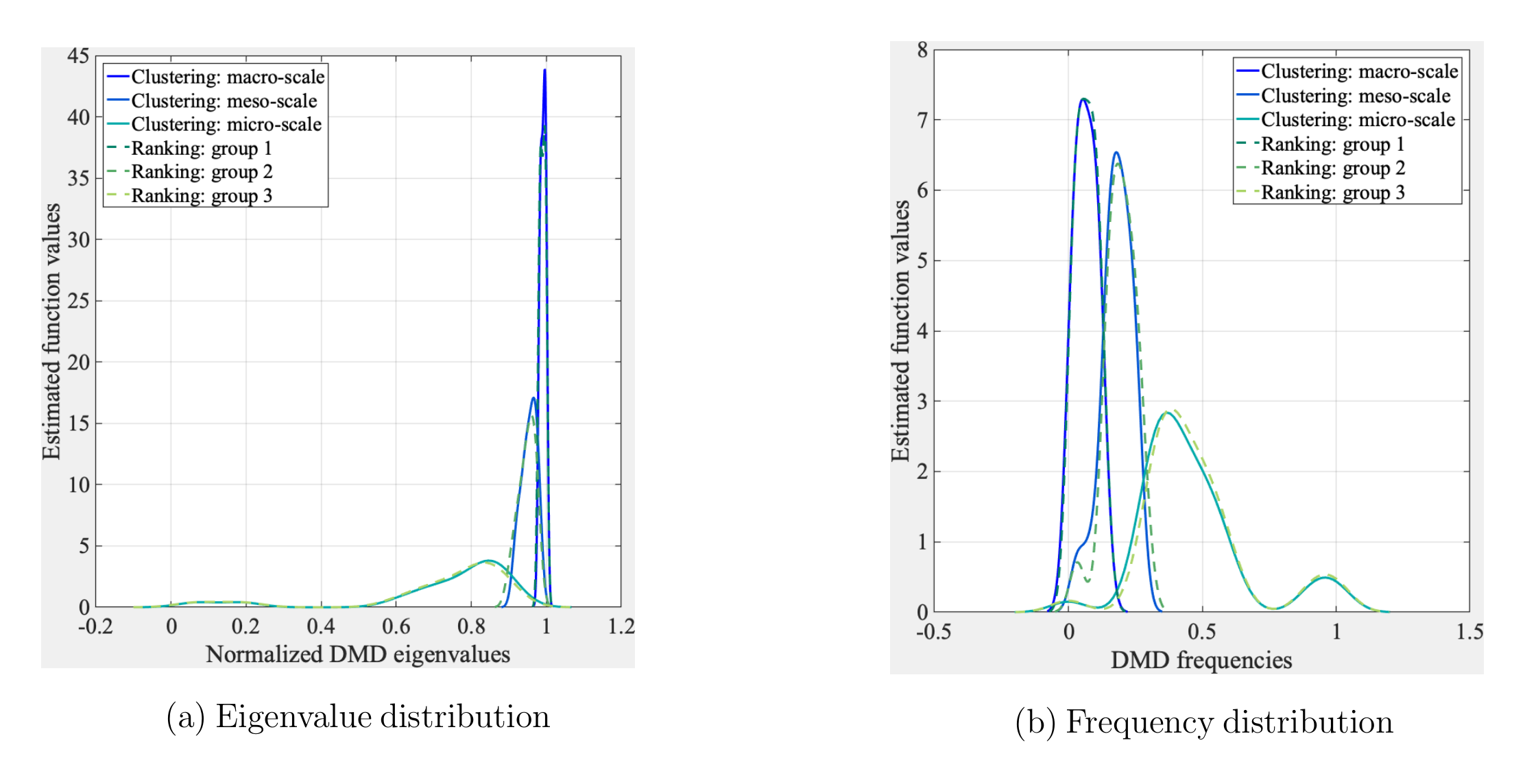}
\caption{Distributions of clustered eigenvalues and frequencies.} 
\label{fig: 12}
\end{figure}

\textbf{\textsl{Clustering results}} Because the decomposition process generates $125$ eigenmodes, an unsupervised machine learning method, i.e., clustering is adopted to group DMD modes of similar properties. Specifically, the k-means clustering scheme has been implemented on account of the fact that it can control the number of clusters \cite{arthur2007k}. We initialized the cluster number $N_{cluster} = 3$, normalized the $125$ DMD eigenmodes, and specified the distance metric involved in the cluster analysis to the correlation type. As a consequence, there are three types of DMD eigenmodes categorized by the scale of spatial patterns, i.e. macro-scale, meso- scale, and micro-scale. On the other hand, we generally ranked the DMD eigenmodes by the real part of the eigenvalue in descending order. And $125$ ranked DMD eigenmodes were then classified into three groups, where group $1$ is $\{ \phi_1, \dots, \phi_{42} \}$, group $2$ is $\{ \phi_{43}, \dots, \phi_{84} \}$, and group $3$ is $\{ \phi_{85}, \dots, \phi_{125} \}$. We approximated the distribution of DMD eigenvalues and frequencies associated with each cluster/group utilizing a normal kernel smoothing function. In \cref{fig: 12}, it can be observed that the clustering and physical grouping results are highly similar. In particular, macro-scale DMD modes tend to have larger eigenvalues while the normalized eigenvalues of micro-scale DMD modes spread over the entire interval. Such finding indicates macro-scale DMD modes contain more energy. In light of the oscillation frequency, macro-scale DMD modes have a relatively lower frequency range compared to the other two types. This observation agrees with the experimental finding on the turbulence of different length scales, where frequencies of large eddies are much smaller than small eddies \cite{tennekes1972first, holmes2012turbulence}.

\begin{figure}[H]
\centering
\includegraphics[width=1.0\textwidth]{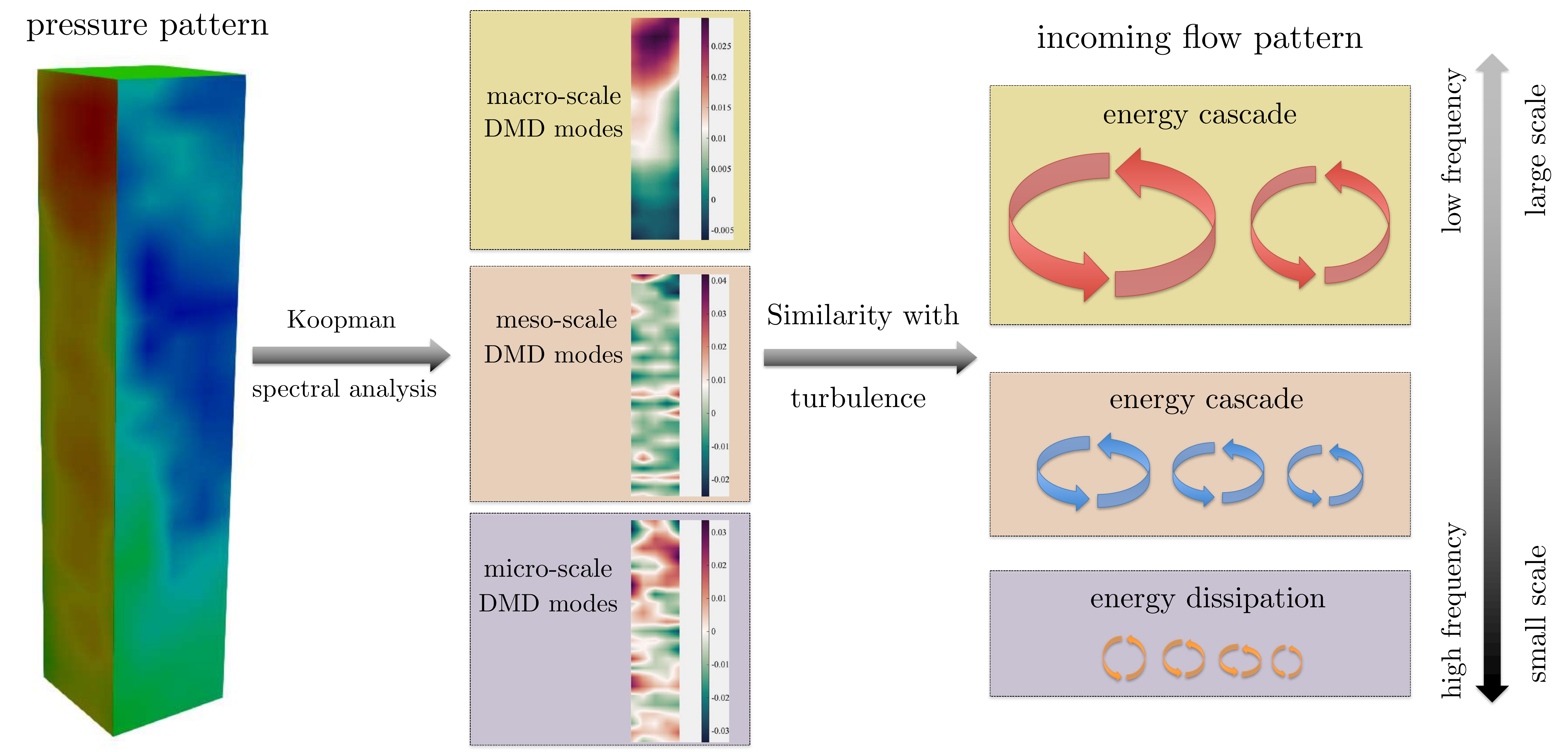}
\caption{Similarities of clustered DMD modes with turbulence energy distribution.} 
\label{fig: 13}
\end{figure}

\textbf{\textsl{Energy cascade}} The clustering results indicate there exists a link between identified DMD modes and turbulent eddies formed in the surrounding wind environment. In this study, interest is in inferring the physical connections between clustered DMD modes and turbulent dynamics generated by the incoming winds. For example, it is known that the aerodynamic fluctuating loads acting on the windward face of the prism model are dominated by the approach flow turbulence. And aerodynamic loads acting on the two side and leeward faces are closely related to the wake dynamics including vortex shedding, which in turn is influenced by the turbulence in the approach flow \cite{vickery1966fluctuating, lee1975effect, aubry1988dynamics}. These experimental findings show the similarity of turbulence properties with the aerodynamic loads represented by combining the DMD modes and their time coefficients. To understand how turbulence related to the aerodynamics pressure field around prisms, we focus on explaining the energy transfer by investigating the dynamical evolution of each DMD mode. In particular, the dynamical evolution process is similar to the energy transferring mechanism in turbulence, which includes energy cascade, neutrally steady, and inverse cascade \cite{leonard1975energy}. Furthermore, it should be noted that the spatial patterns in the pressure field are similar to the distribution of length scales (eddy sizes) in the incoming turbulent flow. \cref{fig: 13} schematically highlights these observations.

\begin{figure}[H]
\includegraphics[width=1.0\textwidth]{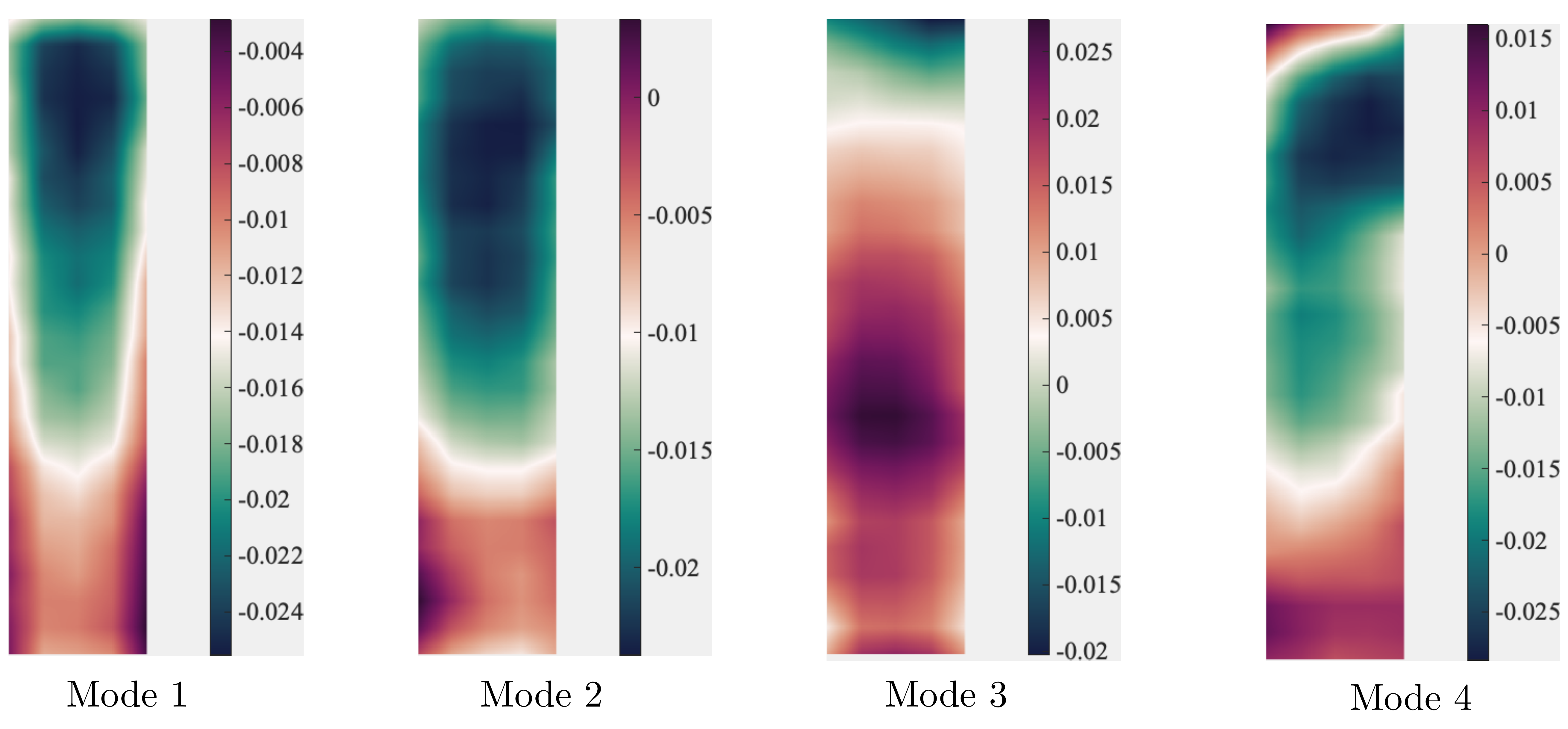}
\caption{Macro-scale DMD modes.} 
\label{fig: f14}
\end{figure}

\textbf{\textsl{Macro-scale DMD modes: linkage to POD modes}} \cref{fig: f14} presents the first four DMD modes in windward scenario. A general view of the resulting DMD modes shows that coherent structures at macro-scale provide a global sketch of the aerodynamic loads. Specifically, DMD modes in the windward scenario are in good agreement with the force pattern reasoning the overturning moment. Locations of local minima and maxima serve as scaled features of lateral forces. The distinct separation of minimum and maximum values indicate the aerodynamic moment on the windward surface computed using these horizontal forces directly contribute to the overall overturning moment. Very similar results have been found in the POD modes \cite{kareem1984pressure, tamura1999proper, carassale2011statistical}.

\begin{figure}[H]
\includegraphics[width=1.0\textwidth]{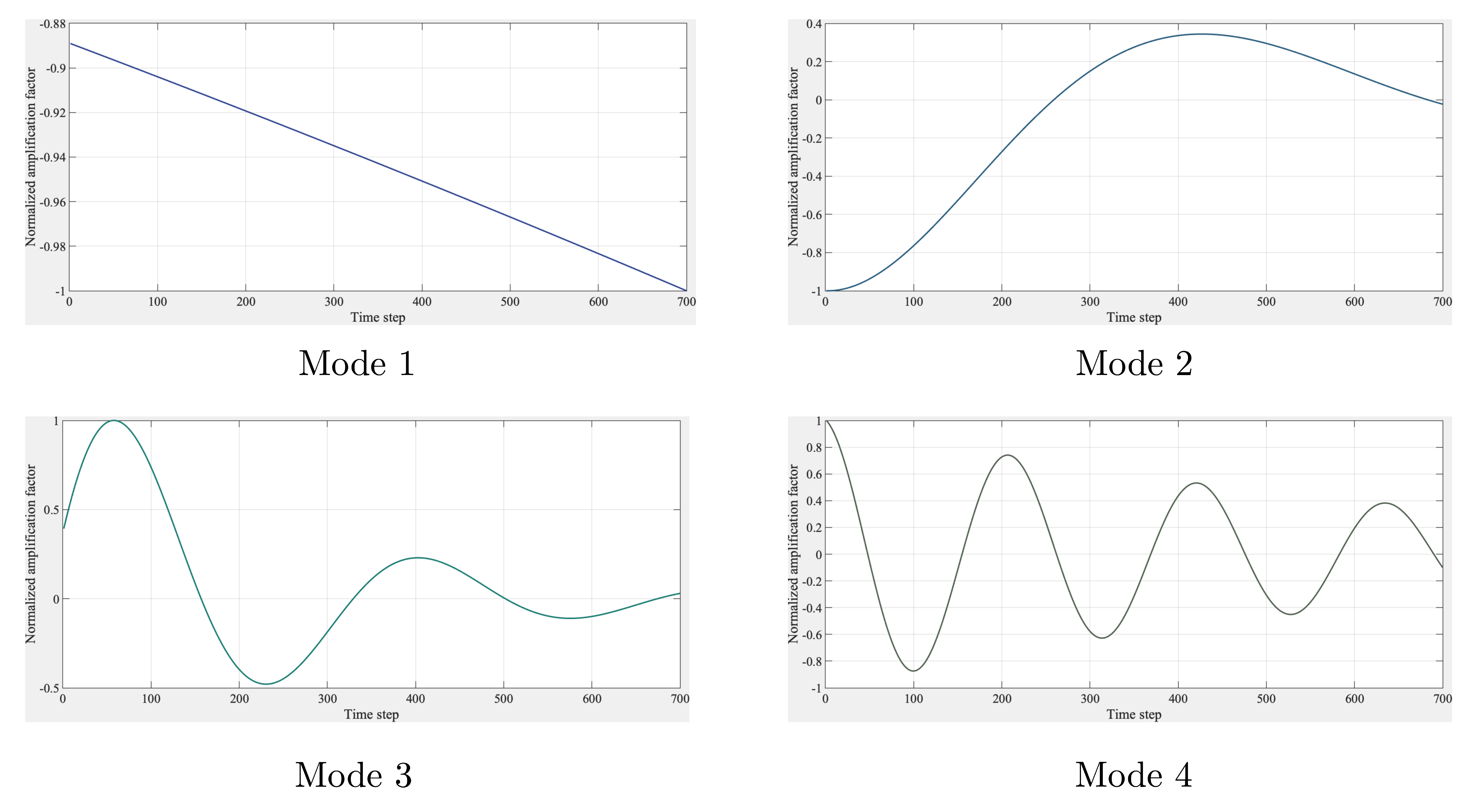}
\caption{Dynamical evolution of selected Macro-scale DMD modes.} 
\label{fig: f15}
\end{figure}

\textbf{\textsl{Temporal evolution of macro-scale DMD modes: linkage to mean flow field}} Because macro-scale DMD modes are more energetic than other types of modes, they are usually assumed to be connected with, to some extent, a collection of flow mechanisms such as the mean flow field around the object \cite{baker2000aspects}. However, one should be aware that a particular DMD mode may be simultaneously related to a few flow mechanisms. In this paper, we are mainly concerned with the mean flow field when analyzing the macro-scale DMD modes. Overall, there are two types of evolution pattern (See \cref{fig: f15} and the supplementary material for more information). First, the absolute amplitude of the first DMD mode appears to grow in all scenarios. This observation conforms to the physical principles of the energy injection concerning stationary boundary layer flows. Specifically, the continually incoming wind constitutes the mean energy injection that balances the mean energy dissipation \cite{holmes2012turbulence}. Second, other DMD modes mostly exhibit decay trend in time. This is a strong resemblance to the turbulent energy cascade, where turbulent eddies of a wide range of scales pass energy successively from larger eddies to the smaller ones \cite{leonard1975energy}. When wind flows over a prism, macro-scale spatial patterns reflect the large-scale fluid motions involved in the approaching wind. Observing at time $t$, the energy associated with these fluid motions is gradually dissipated by the viscosity, resulting in a decaying temporal behavior at $t+\Delta$. And macro-scale coherent structures decompose to a smaller scale (meso or micro). Hence, analyzing the time coefficients of macro-scale DMD modes offer evolution dynamics regarding the break down of fluid flow.

\begin{figure}[H]
\includegraphics[width=1.0\textwidth]{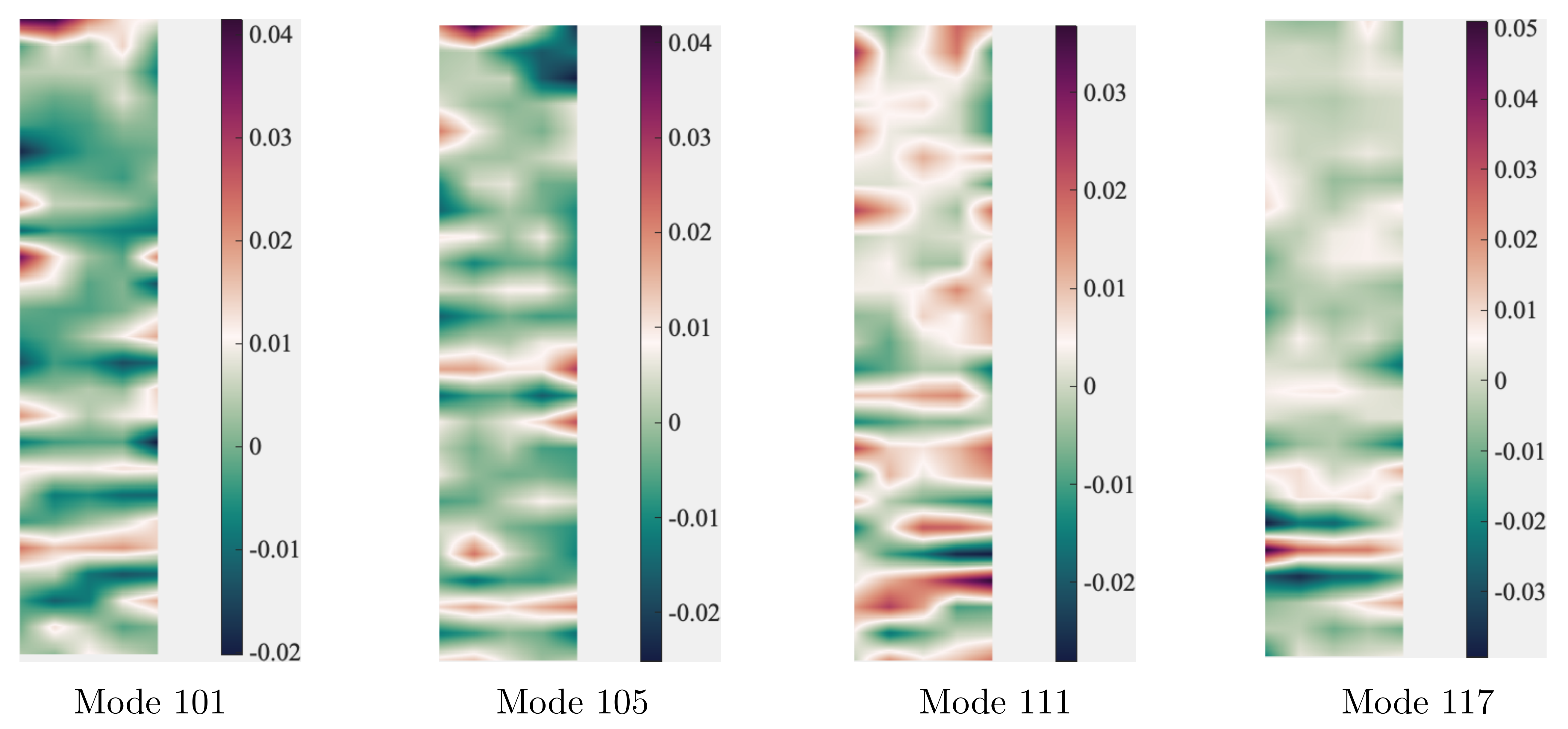}
\caption{Micro-scale DMD modes.} 
\label{fig: f16}
\end{figure}

\textbf{\textsl{Micro-scale DMD modes: linkage to small eddies}} Aerodynamic forces are usually separated into mean and fluctuating components \cite{cermak1976aerodynamics, kareem2008numerical}. And the fluctuating component related to the magnitude of pressure perturbations is directly affected by the nearby small eddies. According to the Kolmogorov length scale theory \cite{frisch1995turbulence}, small eddies tend to have high frequencies, causing turbulence to be isotropic and homogeneous at a local scale. In a generic sense, small eddies represent instabilities of local interactions between different fluid motions. In \cref{fig: f16}, four micro-scale DMD modes from the windward scenario were selected to illustrate such pressure perturbation. Two properties of the resulting DMD modes should be noted. First, pressure perturbations are randomly distributed on each surface. Second, the active areas are relatively small and sparsely scattered. These two properties conform to the characteristics of small eddies (randomness and size) \cite{tennekes1972first, frisch1995turbulence}.

\begin{figure}[H]
\includegraphics[width=1.0\textwidth]{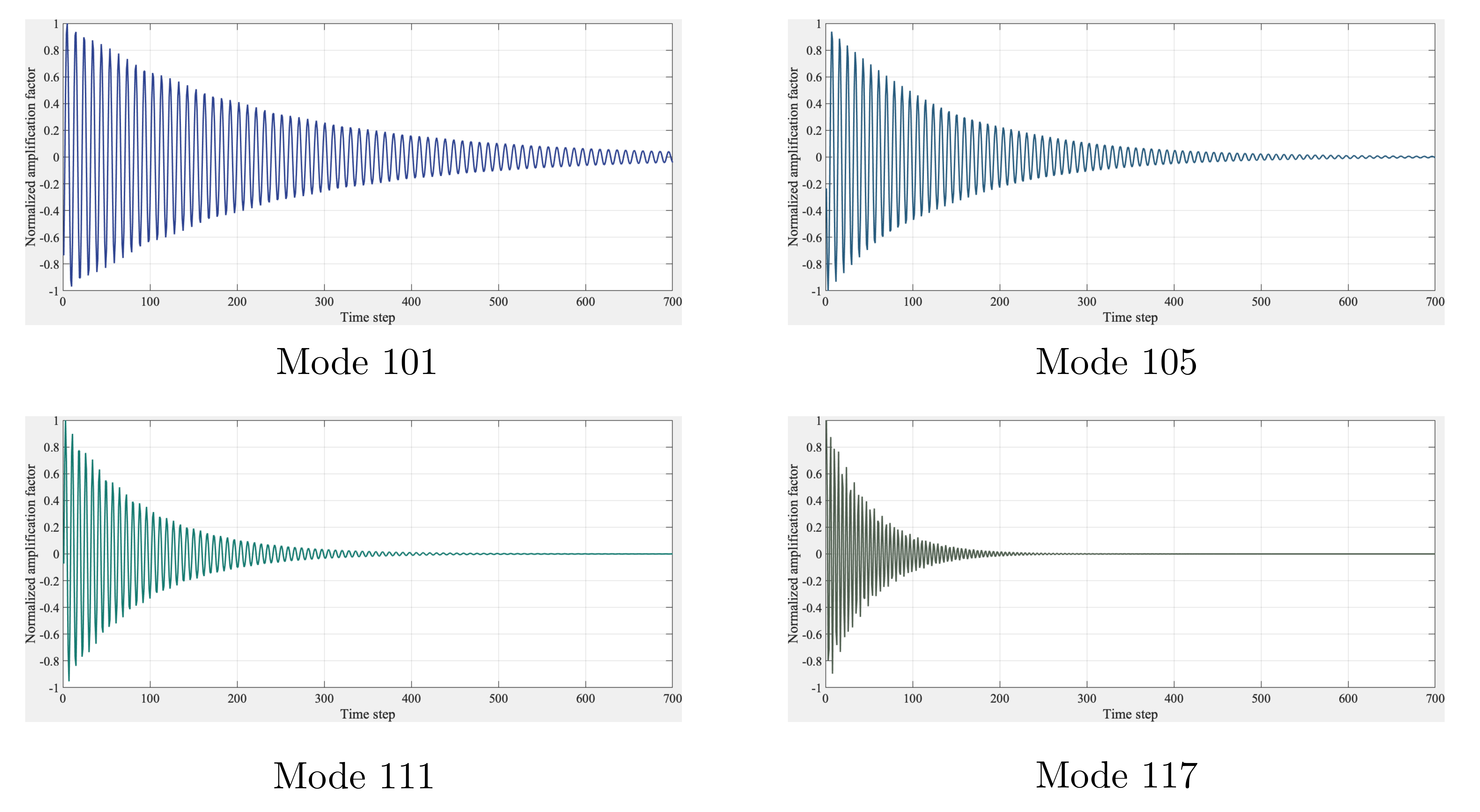}
\caption{Dynamical evolution of selected micro-scale DMD modes.} 
\label{fig: f17}
\end{figure}

\textbf{\textsl{Temporal evolution of micro-scale DMD modes: linkage to energy dissipation}} From a physical standpoint, micro-scale DMD modes do not necessarily represent any specific flow patterns regarding the surrounding flow motions. In \cite{baker2000aspects}, it was pointed that the least energetic modes may contain information on the interaction between different mechanisms instead of actually representing one of them. The conclusion was drawn based on the POD analysis results, where temporal behavior of identified eigenmodes are hard to interpret. In \cref{fig: f17}, we summarize the time coefficients of selected micro-scale DMD modes. It can be clearly seen that all the selected micro-scale DMD modes exhibit decay form. In particular, the amplitude of some modes decays exponentially, for instance, mode $111$ of the windward surface. This reflects the turbulent energy cascade process where large eddies generated by the mean flow cascades down to small eddies, and the kinetic energy carried by small eddies ultimately dissipated to heat in the pressure field \cite{leonard1975energy}.

\begin{figure}[H]
\includegraphics[width=0.9\textwidth]{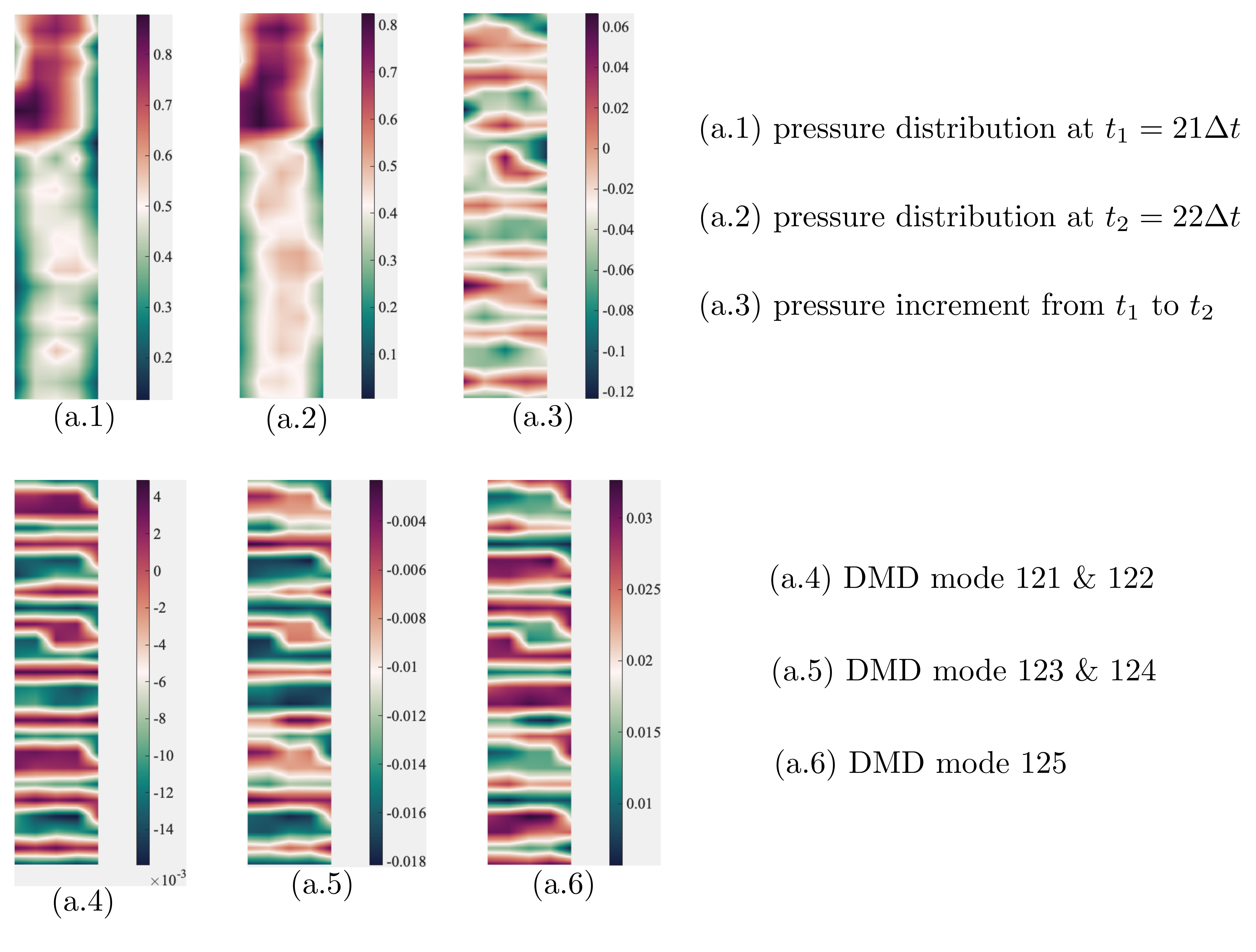}
\caption{Illustration of the connection between micro-scale DMD modes and instantaneous increment of wind pressures.} 
\label{fig: f18}
\end{figure}

\textbf{\textsl{Stripe pattern: linkage to instantaneous pressure increment}} In light of micro-scale DMD modes, there is a specific pattern that is detected on all surfaces of the prism. We call it the stripe pattern because the computed spatial distribution indicates the surface is divided into a few bands. Pressure values of two adjacent bands come from different ranges, thus resembling the stripe pattern observed in nature. To understand these stripe patterns, we examined the evolution of the pressure distribution, finding such micro-scale pattern is closely related to the pressure increment between instantaneous spatial pressure distribution. \cref{fig: f18}, the first row shows the pressure distribution at two consecutive time instances $t_1$ and $t_2$ as well as the instantaneous pressure increment between $p(t_1)$ and $p(t_2)$. The second row displays the identified stripe pattern.

\begin{figure}[H]
\includegraphics[width=1.0\textwidth]{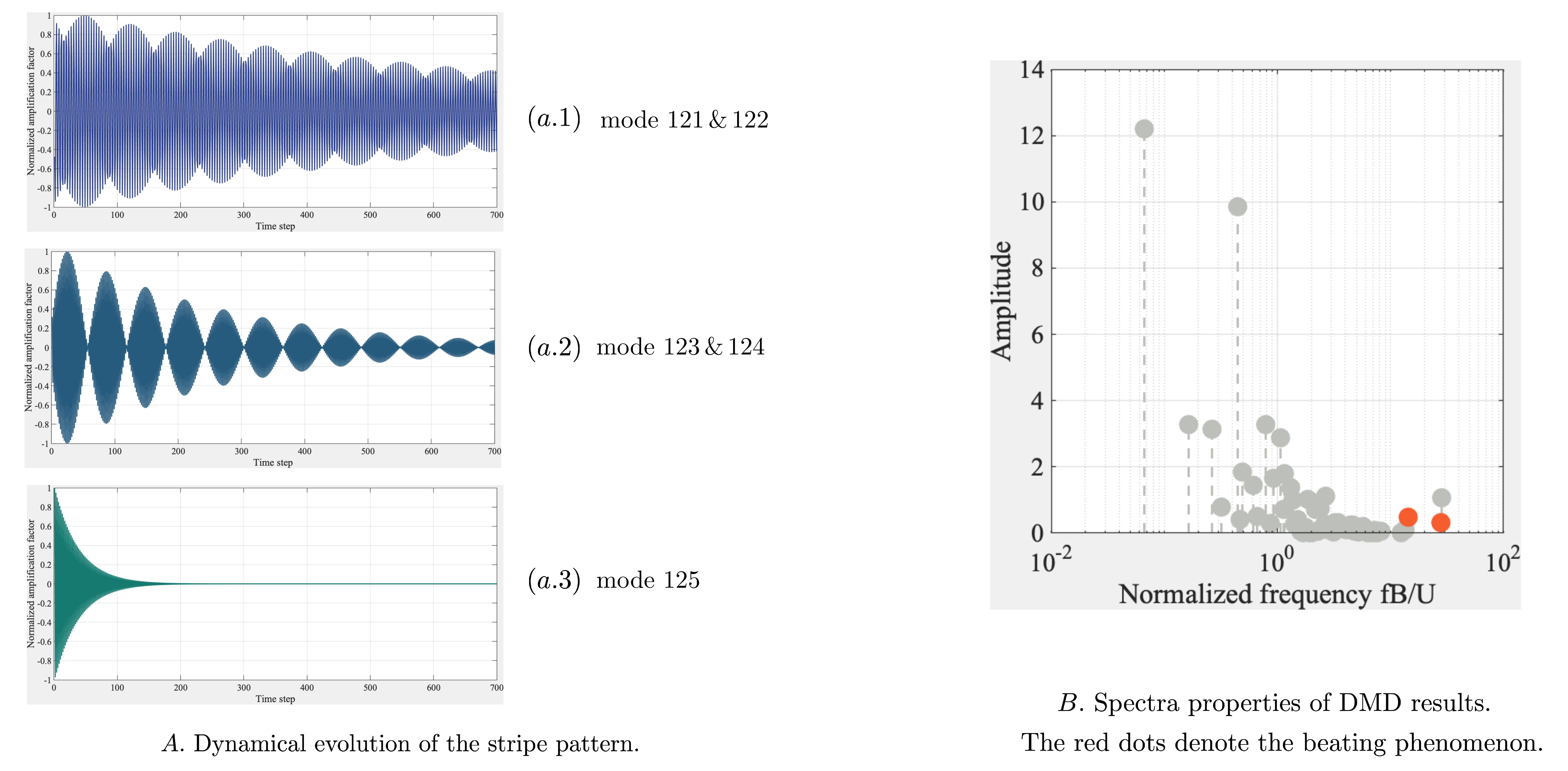}
\caption{Properties of the stripe pattern.}
\label{fig: f19}
\end{figure}

\textbf{\textsl{Temporal evolution of the stripe pattern: linkage to the beating phenomenon}} Eddies of different sizes have their corresponding locations in a frequency spectrum. In \cref{fig: f18}, sizes of the stripe patterns are structurally similar, indicating the size of the surrounding eddies are very close. Consequently, the natural frequencies of these turbulent fluid motions interacting with the prism are close. These factors naturally fulfill the precondition of the beating phenomenon, which in principle is a constructive interference pattern between two signal having almost identical frequencies \cite{yalla2001beat}. The decomposition results describing the dynamical evolution of each stripe pattern mode are given in \cref{fig: f19}. (A). The identified DMD modes oscillate up and down and the amplitudes exhibit a decaying pattern in a general sense. This is due to the fact that small eddies are gradually damped out by viscosity. Moreover, the beat phenomenon that are highlighted by the red dots only happens in the high-frequency region (See \cref{fig: f19}. (B)). An intuitive explanation is based on the definition of the beating phenomenon, which not only requires the existence of two fluid motions sharing very close frequencies but also expects these two eddies are closely located in space. However, lower frequencies are associated with larger fluid motions that are energetically unstable. This precludes the existence of the beating phenomenon in the low-frequency region.

\begin{figure}[H]
\includegraphics[width=1.0\textwidth]{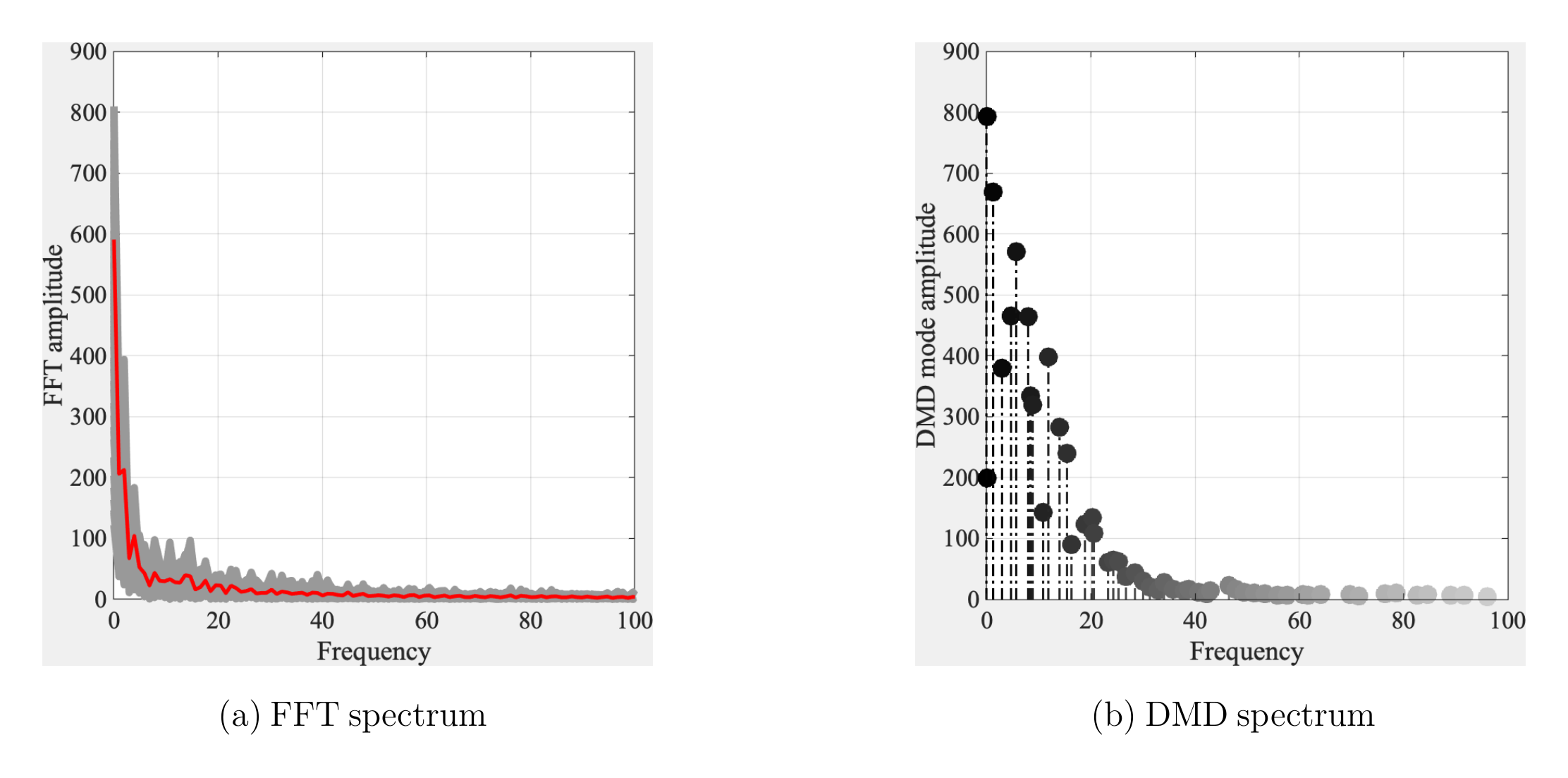}
\caption{DFT and DMD spectrum. In the DFT spectrum representations, the gray band represent the discrete Fourier transform (DFT) results of $125$ time-ordered sequences and the red line denotes their mean.}
\label{fig: f20}
\end{figure}

\textbf{\textsl{Description in discrete Fourier domain}} Chen, K.K et al. pointed the DMD method provides identical results to the temporal discrete Fourier transform (DFT) if the input data is linearly independent and zero centered \cite{chen2012variants}. Despite the fact that our aerodynamic pressure data is highly nonlinear and the mean subtraction is not carried out, DMD spectrum was found to provide useful information regarding the pressure dynamics from the frequency perspective \cite{tu2013dynamic, kutz2016dynamic}. We studied the mode amplitude as a function of frequency, which can be computed from the imaginary part of the DMD eigenvalues. Such spectrum captures the magnitude and phase of pressure dynamics through a linear combination of eigenvectors, where every eigenvector grows or decays at a specified frequency according to its associated eigenvalue. For comparison, we computed the amplitude of oscillations at a wide range of frequencies by means of the Fourier transform, where the temporal DFT model is essentially a superposition of a series of harmonic oscillators. \cref{fig: f20} shows that the DMD spectrum closely resembles the FFT spectrum and the proposed augmented EDMD algorithm can capture the leading amplitudes of the power spectrum computed by the FFT. The difference between DMD and FFT spectrum comes from the algorithm. In particular, FFT spectrum is computed independently using the collected time series data recorded via each pressure tap while every single point in the DMD spectrum denotes a certain coherent structure covering $125$ pressure taps over each surface.

\textbf{\textsl{Review of the temporal evolution}} As wind passes by a building, it results in the formation of Karman vortex shedding. As these vortices are advected by the mean flow, they result in interactive periodic pressure signatures on the building side faces \cite{lee1975effect}. Few attempts have been made to connect identified eigenmodes to the convection of the vortices because of the spatiotemporal nature of these recurrent patterns. In \cite{carassale2012analysis}, the effects of the advection of vortices on the pressure patterns have been identified by repeated operations of digital filter and Hilbert transform of the convolutive data. In this context, the essence of the DMD lies at the use of a paired time-shifted matrices in the computation of eigenmodes, which naturally integrates the evolution dynamics into the decomposition process. In addition, the proposed augmented DMD algorithm (\cref{alg: EDMD}) exploits the delaying coordinates, serving as an alternative to the convolution operation adopted in \cite{carassale2012analysis}.

\begin{figure}[H]
\includegraphics[width=1.0\textwidth]{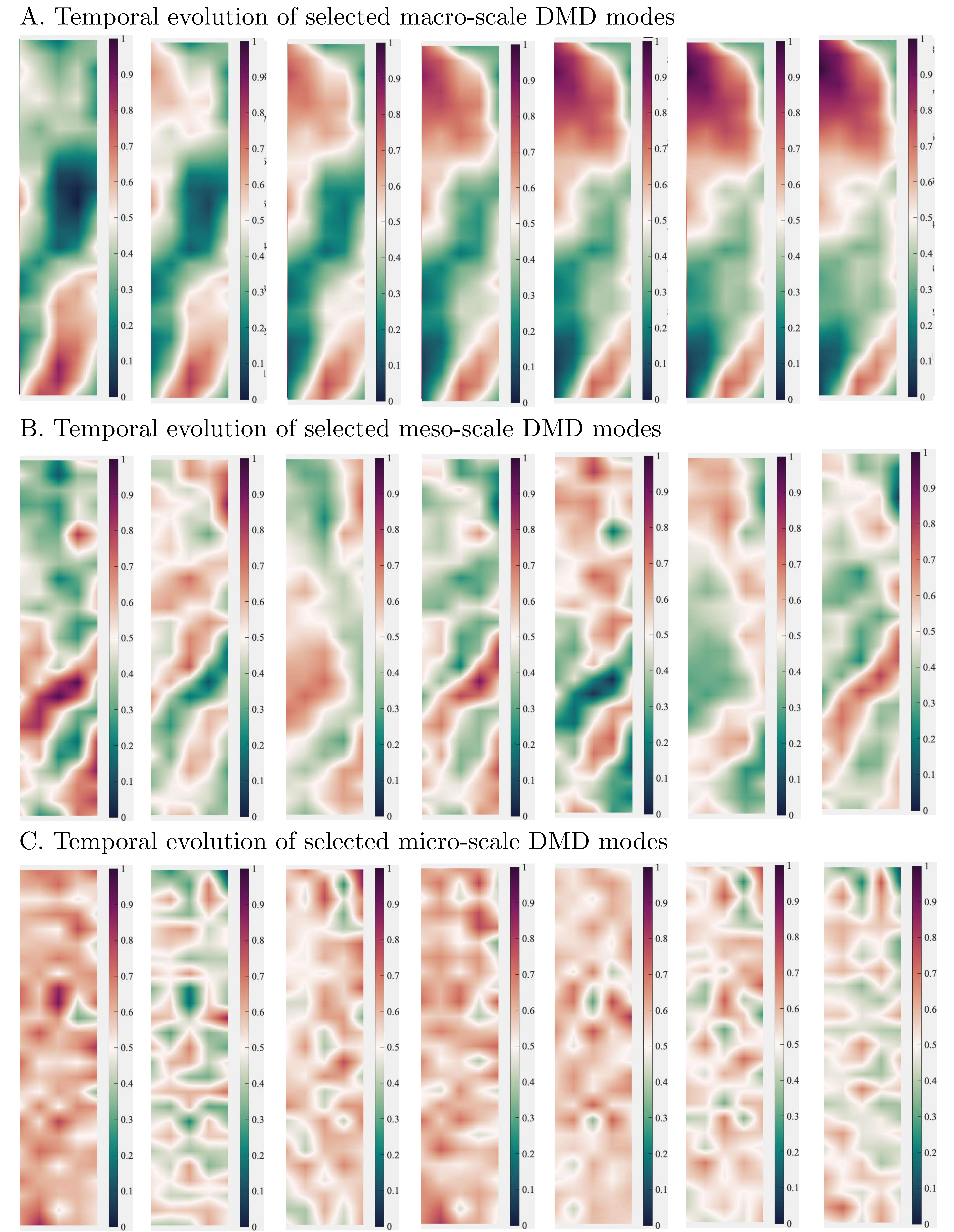}
\caption{Temporal evolution of selected DMD modes of the leftside.} 
\label{fig: f21}
\end{figure}

\cref{fig: f21} shows the temporal evolution of the pressure field at different scales. The time between snapshots equals the sampling period of the measured data $10 \Delta t$, $5 \Delta t$, and $\Delta t$ for the macro-, meso-, and micro-scale modes respectively. A total length of $7$ sequences are presented and a better animated illustration of the dynamical pressure patterns is available at \url{https://xihaier.github.io/}. The macro-scale sequences reported in \cref{fig: f21} clearly shows the development process of the horseshoe vortex and the meso-scale captures the recurrent Karman vortex shedding. Meanwhile, the results confirm the spatial-temporal evolution of selected macro-scale DMD modes can capture the distribution of overall pressure dynamics. Developed at lower frequencies, macro-scale DMD modes serve as the base of aerodynamic pressure fields from an energy perspective. In the case of meso/micro modes, we can observe more local fluctuations with relatively higher frequencies and lower spatial correlation.

\section{Concluding remarks}
\label{sec7}
This study investigates the dynamics of random pressure fields over bluff bodies with an application to pressure field over a prism immersed in a turbulent boundary layer. An operator-theoretic approach is introduced for the aerodynamic characterization using limited wind tunnel test data. First, the highly nonlinear pressure dynamics is linearized in the entire basin by an infinite-dimensional linear operator, i.e. the Koopman operator. Subsequently, spectral analysis of the Koopman operator provides a spatio-temporal characterization of the pressure data. To compute the Koopman eigen-elements, an augmented dynamic mode decomposition method is proposed. The augmentation is achieved by the use of the Takens’s embedding theorem, where time delay coordinates are considered. As a result, the computed DMD representation is spatio-temporally orthogonal where each identified coherent structure is assigned to a specific frequency and a corresponding temporal growth/decay.

To demonstrate the unique feature of the proposed decomposition method, a comparative study between classical POD and proposed augmented DMD is carried out. Unlike POD, which tends to mix the temporal frequencies in the identified modes, DMD provides single-frequency eigenmodes. Furthermore, it is found that the computed DMD modes are closely connected to the wind field approaching the prism. Specifically, similar to the turbulence cascade, macro-scale DMD modes break up and lead to the energy cascade, and micro-scale DMD modes are related to the instantaneous pressure increments and small eddy dynamics.

Future directions regarding applying the Koopman operator to learn similar nonlinear dynamical systems include: (1) Extension of the current work to the extreme learning condition where available data is merely a sequence of a scalar, and (2) Moving from the current Laplacian determinism to learning the intrinsic orderliness governing chaotic system.

\section*{Acknowledgments}
This work was supported in part by the National Science Foundation (NSF) under Grant No. 1562244. This support is gratefully acknowledged. The authors also acknowledge gratefully the use of Tokyo Polytechnic University aerodynamic database. The codes and data used in this work will be made available at \url{https://xihaier.github.io/} upon publication of this manuscript.

\bibliographystyle{elsarticle-num}
\bibliography{sample}

\newpage
\appendix
\section{Appendix. Proper orthogonal decomposition}
\label{app1}
This appendix provides a quick summary of the proper orthogonal decomposition (POD) method. Let $\mathcal{D}$ be the input dataset. Under finite energy assumption, $\mathcal{D}$ belongs to the Hilbert space $\mathcal{H} = L^2 \left( \Omega_{\boldsymbol{T}} \times \Omega_{\boldsymbol{x}} \right)$ of Lebesque measurable. Because the tensor space is dense in $L^2 \left( \Omega_{\boldsymbol{T}} \times \Omega_{\boldsymbol{x}} \right)$, the input dataset takes a tensor product form:

\begin{equation}
\label{eq: app1_1}
\mathcal{D} = \mathcal{A} \left( \Omega_{\boldsymbol{T}} \right) \otimes \boldsymbol{\Phi} \left( \Omega_{\boldsymbol{x}} \right)
\end{equation}

where $\mathcal{A} \left( \Omega_{\boldsymbol{T}} \right) \doteq L^2 \left( \Omega_{\boldsymbol{T}} \right)$, $\boldsymbol{\Phi} \left( \Omega_{\boldsymbol{x}} \right) \doteq L^2 \left( \Omega_{\boldsymbol{x}} \right)$, and $\Omega_{\boldsymbol{x}}$ represents the spatial domain. As a result, the spatio-temporal representation of $\mathcal{D}$ is explicitly converted to a bilinear composition form, with the first vector space $\mathcal{A}$ producing the time-dependent coefficients and the second vector space $\boldsymbol{\Phi}$ giving spatial modes. Mathematically speaking, the preceding equation can be equivalently expressed as a linear superposition using a complete set of basis functions according to the Schmidt decomposition theorem:

\begin{equation}
\label{eq: app1_2}
\mathcal{D} = \sum_{j=1}^{N} a_j \left( t \right) \phi_j \left( x \right)
\end{equation}

Unfortunately, $N$ is a considerably large number or infinity in most cases. The essence of POD is to seek an orthogonal projector $\boldsymbol{\upsilon} \left( \boldsymbol{x}, t \right)$ that minimizes the approximation error:

\begin{equation}
\label{eq: app1_3}
\mathcal{J} \left( \boldsymbol{\upsilon} \right) = || \mathcal{D} - \boldsymbol{\upsilon} \left( \boldsymbol{x}, t \right) ||_{n_F}
\end{equation}

with $|| \cdot ||_{n_F}$ denoting the Frobenius norm. In the context of POD, $n_F = 2$, which guarantees the kinetic energy interpretation of its optimized modes $\phi_j, j=1, 2, \dots, N$, and the projector $\boldsymbol{\upsilon} \left( \boldsymbol{x}, t \right)$ can be directly constructed through utilizing a finite set of orthogonal basis functions ($M \ll N$) following the principles stated in \cref{eq: app1_2}:

\begin{equation}
\label{eq: app1_4}
\mathcal{D} \approx \boldsymbol{\upsilon} \left( \boldsymbol{x}, t \right) = \sum_{j=1}^{M} a_j \left( t \right) \phi_j \left( x \right)
\end{equation}

Then, the optimal basis functions can be obtained by maximzing the average projection of $\mathcal{D}$ onto $\boldsymbol{\Phi}$:

\begin{equation}
\label{eq: app1_5}
\argmax \frac{< \mathcal{D}, \boldsymbol{\Phi} \left( \boldsymbol{x} \right) >_{\Omega_{\boldsymbol{x}}}}{< \boldsymbol{\Phi} \left( \boldsymbol{x} \right), \boldsymbol{\Phi} \left( \boldsymbol{x} \right) >_{\Omega_{\boldsymbol{x}}}}
\end{equation}

By applying the calculus of variations, the optimization problem stated in \cref{eq: app1_5} is transformed to a Fredholm eigenvalue setting, where the optimal POD basis is exactly the eigenfunctions in the following integral equation:

\begin{equation}
\label{eq: app1_6}
\int_{\Omega_{\boldsymbol{x}}} R \left( \boldsymbol{x}, \boldsymbol{x}^{'} \right) \phi_j \left( \boldsymbol{x}^{'} \right) d \boldsymbol{x}^{'} = \lambda_j \phi_j \left( \boldsymbol{x} \right)
\end{equation}

where $R \left( \boldsymbol{x}, \boldsymbol{x}^{'} \right)$ is the symmetric spatial correlation tensor that is computed by taking the inner product of the pressure data in $\Omega_{\boldsymbol{x}}$:

\begin{equation}
\label{eq: app1_7}
R \left( \boldsymbol{x}, \boldsymbol{x}^{'} \right) = < \boldsymbol{\upsilon} \left( \boldsymbol{x}, t \right) \otimes \boldsymbol{\upsilon} \left( \boldsymbol{x}^{'}, t \right) >_{\Omega_{\boldsymbol{x}}}
\end{equation}

Two important properties associated with \cref{eq: app1_6} should be observed. First, Hilbert-Schmidt theory assures the eigenvalue problem has an infinitely countable pair of eigenfunctions and eigenvalues $\{ \phi_j, \lambda_j \}$, out of which there exists a sequence of positive eigenvalues $\lambda_1 \geqslant \lambda_2 \geqslant \dots \geqslant 0$. POD modes are ranked according to the eigenvalues of $R \left( \boldsymbol{x}, \boldsymbol{x}^{'} \right) \phi_j \left( \boldsymbol{x}^{'} \right)$ by a descending order. Second, eigenfunctions are orthogonal $\int_{\Omega_{\boldsymbol{x}}} \boldsymbol{\phi}_i \left( \boldsymbol{x} \right) \boldsymbol{\phi}_j \left( \boldsymbol{x} \right) d \boldsymbol{x} = \delta_{ij}$, that is, the POD algorithm seeks the least number of modes for describing $\mathcal{D}$.

\section{Appendix. Evaluation metrics for convergence study}
\label{app2}
We checked the convergence of the decomposition results using both eigenmodes and eigenvalues. The first evaluation metric is a function of identified eigenmodes. It is used to analyze the convergence behavior of the spatial pattern. Consider two datasets $\mathcal{D}^{1} = \{ \boldsymbol{x} \left( t_1 \right), \dots, \boldsymbol{x} \left( t_{m} \right) \}$ and $\mathcal{D}^{2} = \{ \boldsymbol{x} \left( t_1 \right), \dots, \boldsymbol{x} \left( t_{n} \right) \}$, which contain pressure signals of the same prism surface but with a different sampling interval $[t_1, t_{m}]$ and $[t_1, t_{n}]$. Two sets of eigenmodes $\boldsymbol{\Phi}^{m}$ and $\boldsymbol{\Phi}^{n}$ can be accordingly computed by means of either POD or DMD method. To measure their relative difference, a $L_2$ norm based metric is defined:  

\begin{equation}
\label{eq: app2_1}
\kappa_j = < \phi^{m}_j - \phi^{n}_j, \phi^{m}_j - \phi^{n}_j >_{\Omega_{\boldsymbol{x}}}
\end{equation}

In this study, $n = 30000$, serving as the reference point in investigating convergence properties, and $m$ is sampled from the interval $[\Delta t, n \Delta t]$.

Meanwhile, the second evaluation metric is a function defined based on the normalized eigenvalues:

\begin{equation}
\label{eq: app2_2}
\lambda_j^{\dagger} = \frac{\lambda_j}{\sum_{i=1}^{n} \lambda_i \left( m \right)}
\end{equation}

where $\lambda_j^{\dagger}$ primarily examines the temporal convergence properties. It should be further noted that the DMD eigenvalues $\lambda_j^{DMD}$ and eigenmodes $\phi_j^{DMD}$ are complex-valued. For this reason, complex modulus that measures the magnitude of a tensor from the coordination center to its location in the complex plane is utilized:

\begin{equation}
\label{eq: app2_3}
\begin{aligned}
\lambda_j^{DMD} & = \sqrt{\left( \mathcal{R}e \left( \lambda_j^{DMD} \right) \right)^2 + \left( \mathcal{I}m \left( \lambda_j^{DMD} \right) \right)^2} \\ 
\phi_j^{DMD} & = \sqrt{\left( \mathcal{R}e \left( \phi_j^{DMD} \right) \right)^2 + \left( \mathcal{I}m \left( \phi_j^{DMD} \right) \right)^2}
\end{aligned}
\end{equation}


\end{document}